\documentclass[aps,pre,twocolumn,groupedaddress]{revtex4}
\usepackage{epsfig}
\usepackage{graphicx}
\usepackage{dcolumn}
\usepackage{bm}

\begin{document}

\title{Atypical viral dynamics from transport through popular places}

\author
{Pedro D. Manrique$^{1}$, Chen Xu$^{2}$, Pak Ming Hui$^{3}$ and Neil
F. Johnson$^{1}$}
\affiliation{$^{1}$Physics Department, University of Miami, Coral Gables, FL 33126, U.S.A.\\
$^{2}$College of Physics, Optoelectronics and Energy, Soochow University, Suzhou 215006, China\\   
$^{3}$Department of Physics, The Chinese University of Hong Kong, Shatin, Hong Kong}

\date{\today}

\begin{abstract}
The flux of visitors through popular places undoubtedly influences viral spreading -- from H1N1 and Zika viruses spreading through physical spaces such as airports, to rumors and ideas spreading though online spaces such as chatrooms and social media. However there is a lack of understanding of the types of viral dynamics that can result. Here we present a minimal dynamical model which focuses on the time-dependent interplay between the {\em mobility through} and the {\em occupancy of} such spaces. Our generic model permits analytic analysis while producing a rich diversity of infection profiles in terms of their shapes, durations, and intensities. The general features of these theoretical profiles compare well to real-world data of recent social contagion phenomena. 
\end{abstract}

\maketitle
\section{Introduction}
From the spread of pathogens \cite{flu,manore14,stanley1} through places such as airports, schools \cite{children1} and hospitals \cite{onnela}, to the spread of online popularity \cite{boyle,havlin} and rumors through Internet chatrooms \cite{finance,neil} or bulletin boards \cite{sornette1,riley}, the issue of viral spreading through popular places is of prime importance. Many sophisticated epidemiological models have been proposed of viral dynamics  \cite{Keeling,previous,May,Koopman,Murray,cvespignani,schwartz,blasius,dodds,colizza,Havlin2,us,watts,stanley2,3,murase,vesp,scholtes14,barrat,baron,vesp2,kaski,ker,perra,10,5,11more,vesp09,11,gonc,estrada} with a theoretical focus spanning from the well-mixed (i.e. mass-action) limit through to heterogeneous and even dynamically-evolving networks \cite{cvespignani,schwartz,blasius,dodds,Havlin2,11more,5,10,vesp2,baron,barrat,vesp,stanley2,watts,3,murase,us}. There is however a lack of quantitative understanding of how people revisiting a popular place impacts the detailed profiles that emerge from viral spreading (e.g. school, supermarket, airport or online bulletin board). In particular, the interplay between the mobility through such a space and the average occupancy, has not been addressed in any analytically amenable way to our knowledge.

\begin{figure}[tbp]
\begin{center}
\includegraphics[width=0.9\linewidth]{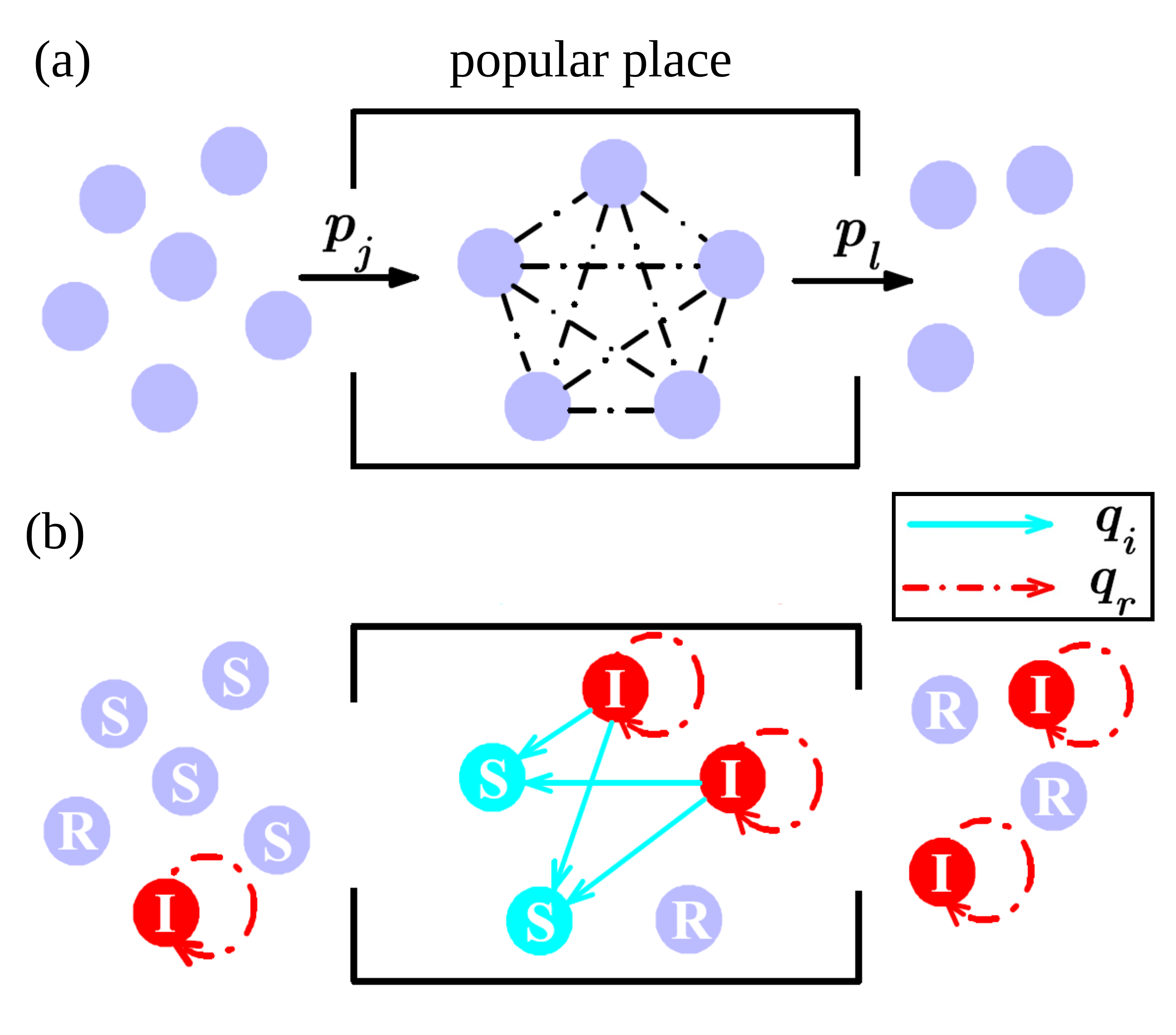}
\end{center}
\caption{(Color online) Schematic diagram showing our model of viral spreading due to revisits to a popular offline or online space. (a) At each timestep, an agent who happens to be outside the popular space, has a probability $p_{j}$ to enter. Meanwhile, an agent who happens to be inside the popular space has a probability 
$p_{l}$ to leave at that timestep. (b) At each timestep, an infected agent who happens to be inside the popular space, has a probability $q_{i}$ to infect any susceptible agent who also happens to be inside the space at that timestep. Also at each timestep, each infected agent, whether inside or outside the space, recovers with probability $q_{r}$ as in standard SIR (Susceptible-Infected-Recovered) models.}
\label{fig1}
\end{figure}

In this paper we present a simple model of co-existing human mobility and infection dynamics. Our model predicts highly non-trivial viral dynamics due to the direct interplay between the mobility through, and the average occupancy of, a generic popular space $G$ (see Fig. \ref{fig1}(a)). In Sec. II we introduce the model and its variants. Section III focuses on the SIR (Susceptible-Infected-Recovered) process. We analyze the co-evolving dynamical equations numerically and analytically. We obtain highly diverse infection profiles $I(t)$ as a function of the mobility and occupancy of the public place. Even when each individual agent spends the same average time in $G$ and has the same average number of contacts, and the average attendance of $G$ is constant, infection profiles arise which are qualitatively different from the well-mixed limit (e.g. resurgent epidemics). We then derive a specific analytic condition involving the mobility and occupancy, for which the co-existing dynamical processes reduce to an effective SIR model with renormalized parameters. More generally, we study the variation in shape of the infection profile by looking at its extensive features such as duration, severity and time-to-peak, and uncover an interesting linearity between the infection probability and the mobility. We use this analysis to compare the outcome of the model with modern-day outbreaks from the social domain, finding good agreement. 

We then compare these results to a broadcast-type infection mechanism, where infection occurs through an individual's presence in the popular space (e.g. viruses on surfaces, or an endemic population of infected mosquitos) as opposed to through another infected person, and hence the infection process does not depend on the other individuals in the popular space. We find that it is only in the specific limit of the infection probability being much greater than the recovery probability, that the results are similar for these two distinct mechanisms. The significant qualitative differences that we otherwise observe, suggest that distinct policies need to be implemented by planners when dealing with infected individuals (e.g. students or travelers) as opposed to infected spaces (e.g. schools or airports). 

In Sec. IV we present results for our model with an SIS (Susceptible-Infected-Susceptible) process. We find that the system is still tractable by means of a set of differential equations. In addition, we uncover a set of conditions under which the system resembles a standard SIS model and hence the evolution of the subpopulation of susceptibles can be obtained analytically. We also employ the same analysis in the corresponding subsection discussing the broadcast mechanism. Sec. V provides the summary and discussion.


\section{Model}
Figure 1(a) illustrates our model of $N$ agents (e.g. people) with access to a popular space $G$. At any given timestep, an agent not in $G$ has a probability $p_{j}$ to join $G$ while somebody inside $G$ has a probability $p_{l}$ to leave. This effectively generates a dynamical group in $G$, with an  occupancy $N_g(t)$ (i.e. number of agents in $G$) which can fluctuate arbitrarily in time. Two useful combinations of $p_{j}$ and $p_{l}$ are: $\gamma_{s}=p_{j}/(p_{j}+p_{l})$ and $\gamma_{m}=2p_{j}p_{l}/(p_{j}+p_{l})$.  Note that $1/\gamma_{m}$ is the average of $1/p_{j}$ and $1/p_{l}$, i.e. $1/\gamma_{m} = \frac{1}{2} (1/p_{j} + 1/p_{l})$.  The mean number of agents in $G$ is $\langle N_{g}(t) \rangle = Np_{j}/(p_{j}+p_{l}) = N\gamma _{s}$. The total number of agents joining and leaving $G$ on average in a timestep is $N\gamma _{m}$, which characterizes the mobility of the agents.  When $p_{j}$ and $p_{l}$ are scaled by a factor $r$, $\gamma_{s}$ and hence $\langle N_{g}(t) \rangle$ remain unchanged while the mobility changes by a factor $r$. Hence varying $\gamma_{m}$ with
fixed $\gamma_{s}$ amounts to changing the mobility while keeping $\langle N_{g}(t) \rangle$ fixed.
Figure 1(b) shows the effect of adding SIR infection dynamics. At any timestep, any infected person within the popular space $G$ can transmit a virus to any susceptible in $G$ with probability $q_{i}$ (i.e., SIR or SIS mechanism). Later we consider another mechanism where there is a constant probability for every person within the popular space to get infected (i.e. broadcast mechanism). In both cases, no transmission can occur from infected individuals outside $G$. By contrast, since recovery is an individual-based phenomenon, infected individuals both inside and outside $G$ have a probability $q_{r}$ to either recover and become immune (i.e. SIR mechanism) or recover and become susceptible (i.e. SIS mechanism).  It is convenient to define an infected individual's contact rate $\lambda=q_{i}/q_{r}$, which is proportional to the basic reproduction rate of an SIR infection in a well-mixed population \cite{Murray}. We note that we can equivalently view the agents in $G$ as instantaneously connected -- hence our model and results can represent $N$ agents in a time-dependent network, or be applied to the common real-world scenario of a social group with time-varying membership \cite{Palla}. Our regime of focus in this paper, in which a popular space has fairly constant occupancy but variable throughput, is of direct relevance  to online social groups in for example multi-player online games, where it is known that these groups (e.g. guilds) have a size that is fairly constant yet a membership that changes rapidly overtime \cite{us2}. We stress however that our model is far more general in that it allows for any rate of change of occupancy and throughput.

\begin{figure}[tbp]
\begin{center}
\includegraphics[width=0.95\linewidth]{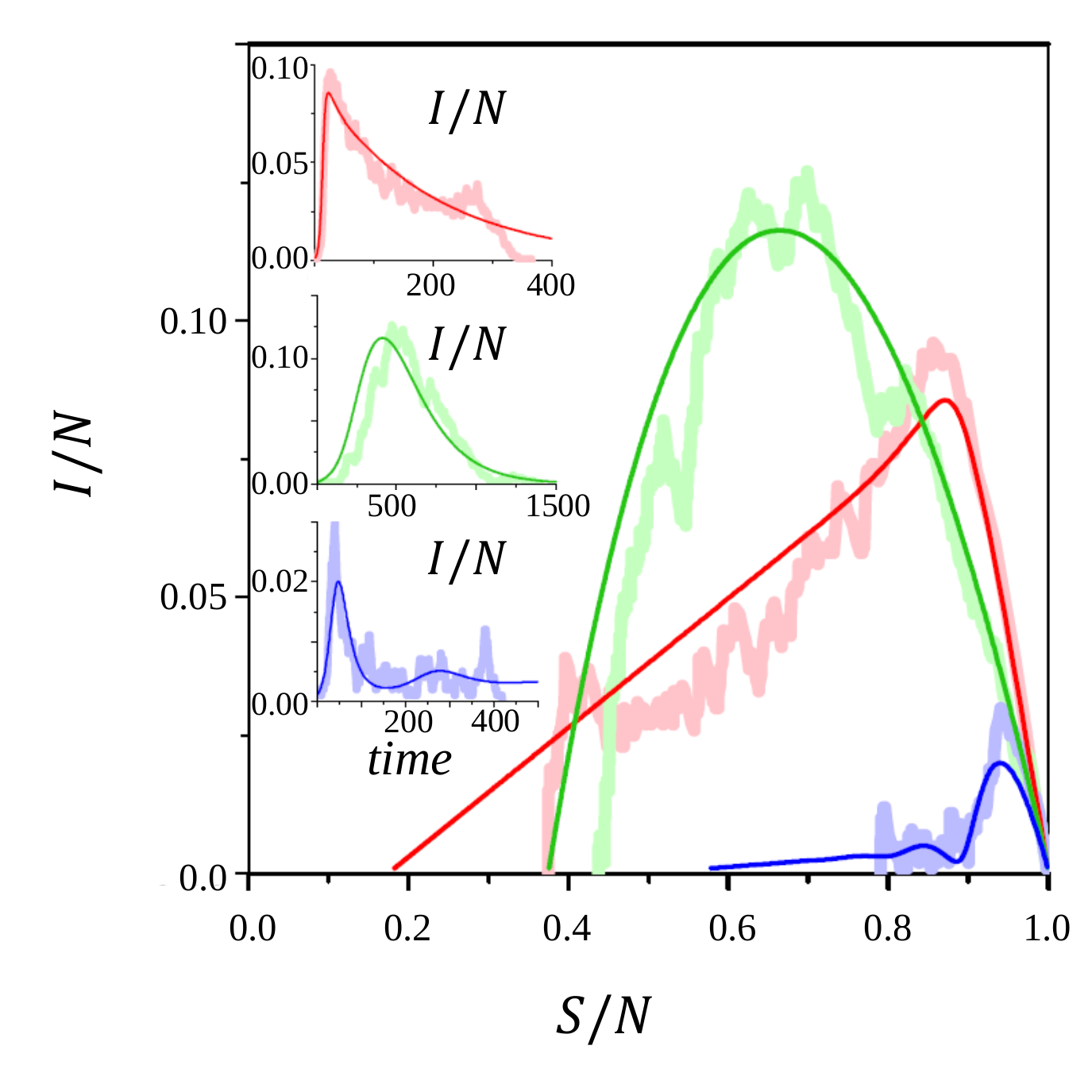}
\end{center}
\caption{(Color online) Trajectories of evolution of the system in the $S$-$I$ space for three different sets of parameter choices.  The rougher and smoother looking curves are obtained by numerical simulation and by integrating the set of differential 
equations (Eq. \ref{SIReqsP2P}) respectively.  The three setups have the same mean number of agents in $G$ given by $N\gamma_{s}$ where $\gamma_{s}=0.1$. Other parameters are: (Red curves) $\gamma_{m}=0.009$, $\lambda=0.1$, and $q_{i}=0.005$; (Green curves) $\gamma_{m}=0.018$, $\lambda=0.1$, and $q_{i}=0.001$; (Blue curves) $\gamma_{m}=0.0018$, $\lambda=0.022$, and $q_{i}=0.002$. Insets show the time dependence of the total system infection level $I(t)$ for each of the cases.}
\label{fig2}
\end{figure}


\section{SIR case}
\subsection{Person-to-person contagion}
\subsubsection{Analysis}
Figure \ref{fig2} summarizes the rich diversity of behaviors which emerge from our model, and the close agreement between individual runs of the numerical simulation and the coupled differential equations that we describe below. The main panel shows the trajectory of $S$ and $I$ values in the system in the $S$-$I$ space. The trajectory starts from the lower right-hand corner, as initially we have $S/N \sim 1$ and $I/N \sim 0$.  The results are in sharp contrast with the SIR model in a well-mixed population in which once $\lambda$ and $I(0)$ are given, the trajectory is fixed \cite{Murray}. For standard SIR in a well-mixed population, if $\lambda$ and $I(0)$ are given, then there will only be one trajectory in the $S$-$I$ space. In the simulations, all agents are initially susceptible and we allow the system to run until the group size in $G$ (i.e. popular space occupancy) reaches its steady-state value $N\gamma _{s}$.  We then randomly pick an agent in $G$ and make it infected. In every subsequent timestep, all the agents first carry out the SIR process followed by the joining or leaving of $G$.  We choose $N=1000$.  The number of recovered agents $R$ at the end of the epidemic reflects the extent of the infection.
We now derive a set of equations for this system.  Since the S$\rightarrow$I process only occurs inside the group (i.e. inside the popular space), we use $S(t)$, $I(t)$, $R(t)$ for the number of susceptible, infected, and recovered agents in the whole system, and $S_{g}(t)$, $I_{g}(t)$, and $R_{g}(t)$ for the corresponding numbers in the space $G$.  The six equations that
describe the dynamics of a SIR process for a single dynamical group (i.e. a single popular space) are as follows, with the subscript $g$ on a variable denoting that variable applies to agents within the space $G$:
\begin{eqnarray}
\frac{dS_{g}}{dt}
&=&-q_{i}S_{g}I_{g}-p_{l}(S_{g}-q_{i}S_{g}I_{g})+p_{j}(S-S_{g}),
\nonumber
\\
\frac{dI_{g}}{dt}
&=&q_{i}S_{g}I_{g}-q_{r}I_{g}-p_{l}(I_{g}+q_{i}S_{g}I_{g}-q_{r}I_{g})\nonumber\\
&&+(1-q_{r})p_{j}(I-I_{g}),\nonumber \\
\frac{dR_{g}}{dt}
&=&q_{r}I_{g}-p_{l}(R_{g}+q_{r}I_{g})+p_{j}((R-R_{g})\nonumber\\
&&+q_{r}(I-I_{g})),\nonumber
\\
\frac{dS}{dt} &=&-q_{i}S_{g}I_{g},\nonumber \\
\frac{dI}{dt} &=&q_{i}S_{g}I_{g}-q_{r}I,\nonumber \\
\frac{dR}{dt} &=&q_{r}I.
 \label{SIReqsP2P}
\end{eqnarray}
The extra terms $-p_{l}q_{i}S_{g}I_{g}$ and $-p_{l}(q_{i}S_{g}I_{g}-q_{r}I_{g})$ as well as the factor $(1-q_{r})$ in Eq. \ref{SIReqsP2P} could in principle be excluded when considering particular real-world applications, depending on the precise details of the discrete time processes involved, i.e. whether one can simultaneously enter or leave the public space while changing infection status. We have checked how the omission of these terms affects the numerical simulations of the equations -- and we find that it makes little difference to the results (less than $10\%$). This makes sense since they represent higher-order interaction terms in the mean-field equations. We choose to retain them in the remainder of the paper, noting that the best implementation of these equations for a particular real-world situation may differ slightly according to the details of how individuals join and leave the public space in question, and the details of whether the infection and recovery processes continue during those dynamical changes. This level of detail is outside the scope of our present paper given that we wish to focus on the discrete-time stochastic results.\\

The insets in Fig. \ref{fig2} show the time-variation of the number of infected agents $I(t)$ for three different sets of parameters that correspond to the same mean number of agents $\langle N_{g} \rangle$ in $G$. Depending on the agents' mobility and infection and recovery probabilities, $I(t)$ (insets) shows qualitatively different behavior including a rapid increase with a gradual drop (red curve), gradual increase and drop in number (green), and an $I(t)$ that shows resurgence (oscillatory $I(t)$) after the initial increase and drop (blue curve).  We stress that these are real oscillatory behaviors, not simply fluctuations.  These oscillations (or more generally, resurgent behavior) also appear in the results obtained from integrating the set of equations (Eq.(\ref{SIReqsP2P})), although the resulting curve is smoother since an average over many runs is implicitly implied by the equations. The resurgence arises from the fresh supply of susceptible and infected agents when new agents join the group.

\begin{figure}
\includegraphics[width=0.8\linewidth]{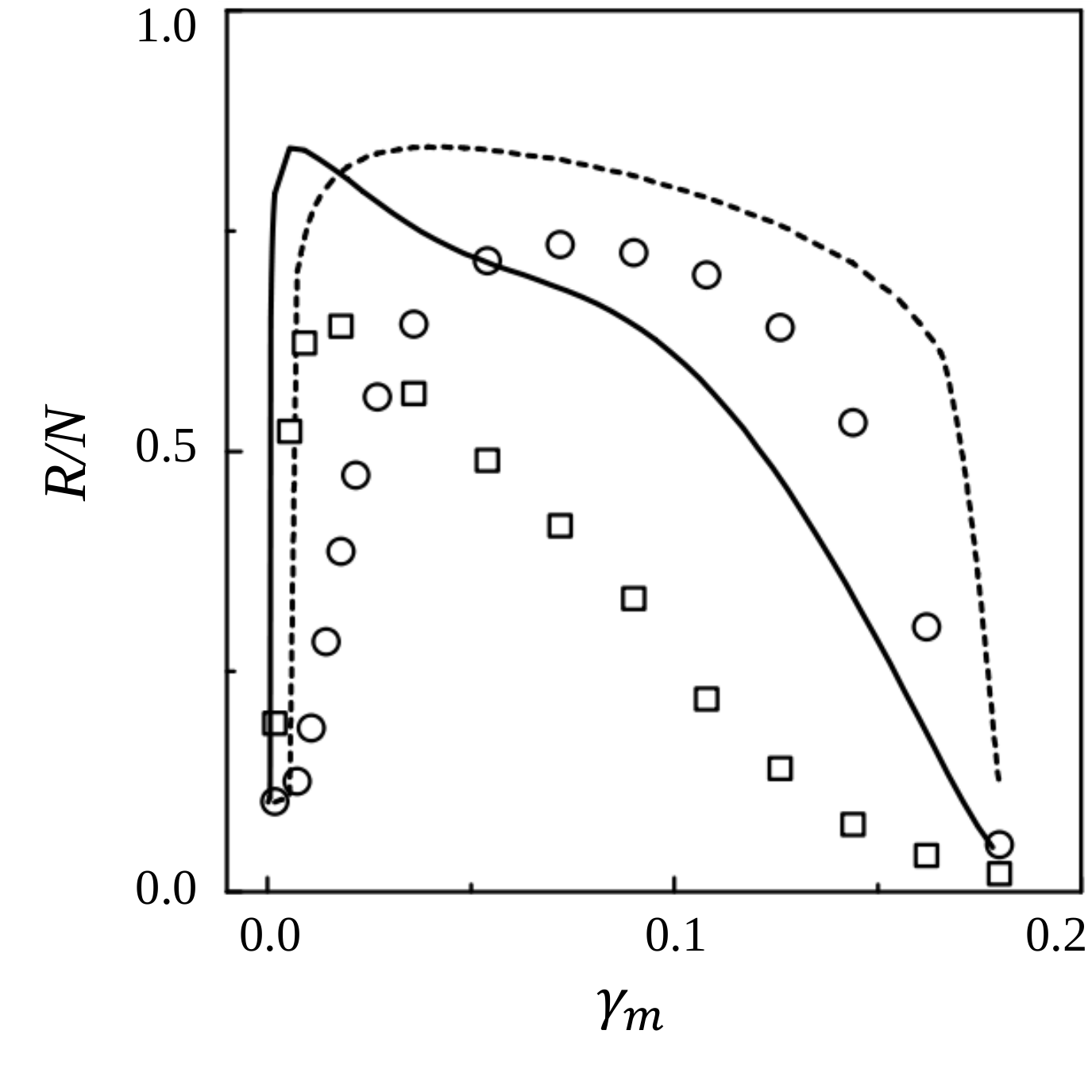}
\caption{Dependence of the final fraction of recovered agents $R/N$ (which is the same as the fraction of agents
who have been infected) on the mobility $\gamma_m$ as obtained by simulations (symbols) and by integrating the set of equations in Eq. \ref{SIReqsP2P} (lines). Results are shown for two systems with the same values of $\gamma_s = 0.1$ and $\lambda = 0.1$ but different infection probabilities. Squares and solid line: $q_i = 0.01$. Circles and dashed line:
$q_i = 0.08$. Simulation results are obtained by averaging over $10^3$ different runs corresponding to
different initializations for a given set of parameters.}
\label{fig3}
\end{figure}

Interestingly, our results show that a large mobility does not necessarily imply more agents become infected and hence that a large $R$ arises in the long time limit. Instead Fig. \ref{fig3} shows the resulting fraction $R/N$ as a function of the mobility $\gamma_m$, for two systems with the same $\gamma_s$ (hence group size) and $\lambda$ (ratio of infection and recovery probabilities) but different infection probabilities. For each case, there is some particular value of $\gamma_m$ that leads to a maximum $R$. The set of equations also gives results that are in qualitative agreement with simulation results. Note that even though the simulations (data points) and equations (lines) do not coincide exactly, the shapes are in reasonable agreement while the results from the equations are consistently higher than the simulation results. A key difference between the simulations and equations is that the number of agents of a certain type is discretized, i.e. $0, 1, 2, 3$ etc. in the simulations. When integrating the differential equations, the associated quantities are taken to be continuous, thus we could have $0<I(t)<1$ when obtaining results using the equations.

\subsubsection{Analytics}
To determine when $I(t)$ will grow initially and create an epidemic, we see from the equation
\begin{equation}
\frac{dI}{dt}=q_{i}S_{g}I_{g}-q_{r}I
\end{equation}
that this can occur when the initial $dI/dt > 0$. At $t=0$, the $S(0)$ initial susceptibles are randomly distributed in that there is no bias in them initially occupying the space $G$ or not. In this case, $S_g(0) = \gamma_s S(0)$ and $I_g (0) = \gamma_s I(0)$. Requiring the right-hand side of the above equation to be greater than zero implies the criterion $\lambda\gamma_{s}^{2}S(0)>1$. We will initialize the infection with one infected agent inside the group, $I_g (0) = 1$ and $S_g (0) = N \gamma_s$. In this case, the criterion for having an initial increase in $I$ is given by $\lambda N \gamma_s > 1$.

Under certain restrictive conditions, it is possible to regard our dynamical model as an effective SIR process in which the effective infection probability is $\gamma_{s}^{2}q_{i}$ and the effective recovery probability is $q_{r}$. The conditions for this to hold are that $p_j + p_l = 1$ (see below) and that the infection probability $q_i$ is sufficiently small so that the number of newly infected agents $q_i S_g (t)I_g (t)$ is less than the number of susceptible $S_g (t)$ in the space $G$. Recalling that the two probabilities are in general treated as independent parameters, we stress that this condition poses an additional restriction on the parameters and hence is not in general true. We can then write the last three equations in Eq. \ref{SIReqsP2P} as:
\begin{eqnarray}
\frac{dS}{dt}&=&-q_{i}S_{g}I_{g}=-q_{i}\left(N\gamma_{s}\frac{S}{N}\right)\left(N\gamma_{s}\frac{I}{N}\right)\nonumber\\&=&-\gamma_{s}^{2}q_{i}SI,\nonumber\\
\frac{dI}{dt}&=&q_{i}S_{g}I_{g}-q_{r}I=q_{i}\left(N\gamma_{s}\frac{S}{N}\right)\left(N\gamma_{s}\frac{I}{N}\right)-q_{r}I\nonumber\\&=&\gamma_{s}^{2}q_{i}SI-q_{r}I,\nonumber\\
\frac{dR}{dt}&=&q_{r}I,
\end{eqnarray}
so that the three equations are the standard SIR equations in a well-mixed population with an effective infection probability of $\gamma_s^{2} q_i$ and an effective recovery probability of $q_r$. Physically, it means that $S_g (t) = N \gamma_s S(t)/N = \gamma_s S(t)$ and $I_g (t) = N \gamma_s I(t)/N =\gamma_{s}I(t)$ for every time step. The system therefore behaves as if all the susceptible and infected agents inside and outside the space $G$ are randomly mixed and may be re-assigned to $G$ at every time step. We now derive the condition $p_j + p_l = 1$ by starting with the discrete time version of $dS_g/dt$:
\begin{eqnarray}
S_{g}(t+1)&=&S_{g}(t)-q_{i}S_{g}(t)I_{g}(t)-p_{l}(S_{g}(t)\nonumber\\&&-q_{i}S_{g}(t)I_{g}(t))+p_{j}(S_{t}-S_{g}(t))\nonumber.
\end{eqnarray}
From $dS/dt = q_i S_g (t)I_g (t)$, we have
\begin{eqnarray}
S(t+1)=S(t)-q_{i}S_{g}(t)I_{g}(t).\nonumber
\end{eqnarray}
The effective dynamical equations are valid if we can write $S_g (t + 1) = \gamma_s S(t + 1)$. Imposing this equality in the above equations, we have
\begin{eqnarray}
(1-p_{j}-p_{l})S_{g}+(p_{j}-\gamma_{s})S-(1-p_{l}-\gamma_{s})q_{i}S_{g}I_{g}=0.\nonumber
\end{eqnarray}
This equality can {\em only} be true at all times if  $p_j + p_l = 1$, for which the coefficients in the three terms then all vanish. Though this condition is restrictive, it is interesting in that it says that the system behaves as a well-mixed SIR model when the probability for the agents outside the space $G$ \textit{not to join} the space (i.e. $(1-p_j )$) is equal to the probability of those inside the space $G$ to leave. Equivalently, the probability for agents inside the space $G$ to stay (i.e., $(1-p_l )$) must be equal to the probability of those outside the space $G$ to join. Under this condition, we no longer need two parameters and a single $p_j$ is sufficient: hence $\gamma_s = p_j$ and the mean group size is $N p_j$. The dynamics of the model then become: An agent outside the space $G$ has a probability $p_j$ to join $G$ and every agent inside $G$ has a probability $(1-p_j)$ to leave.
\begin{figure*}
\includegraphics[width=0.85\linewidth]{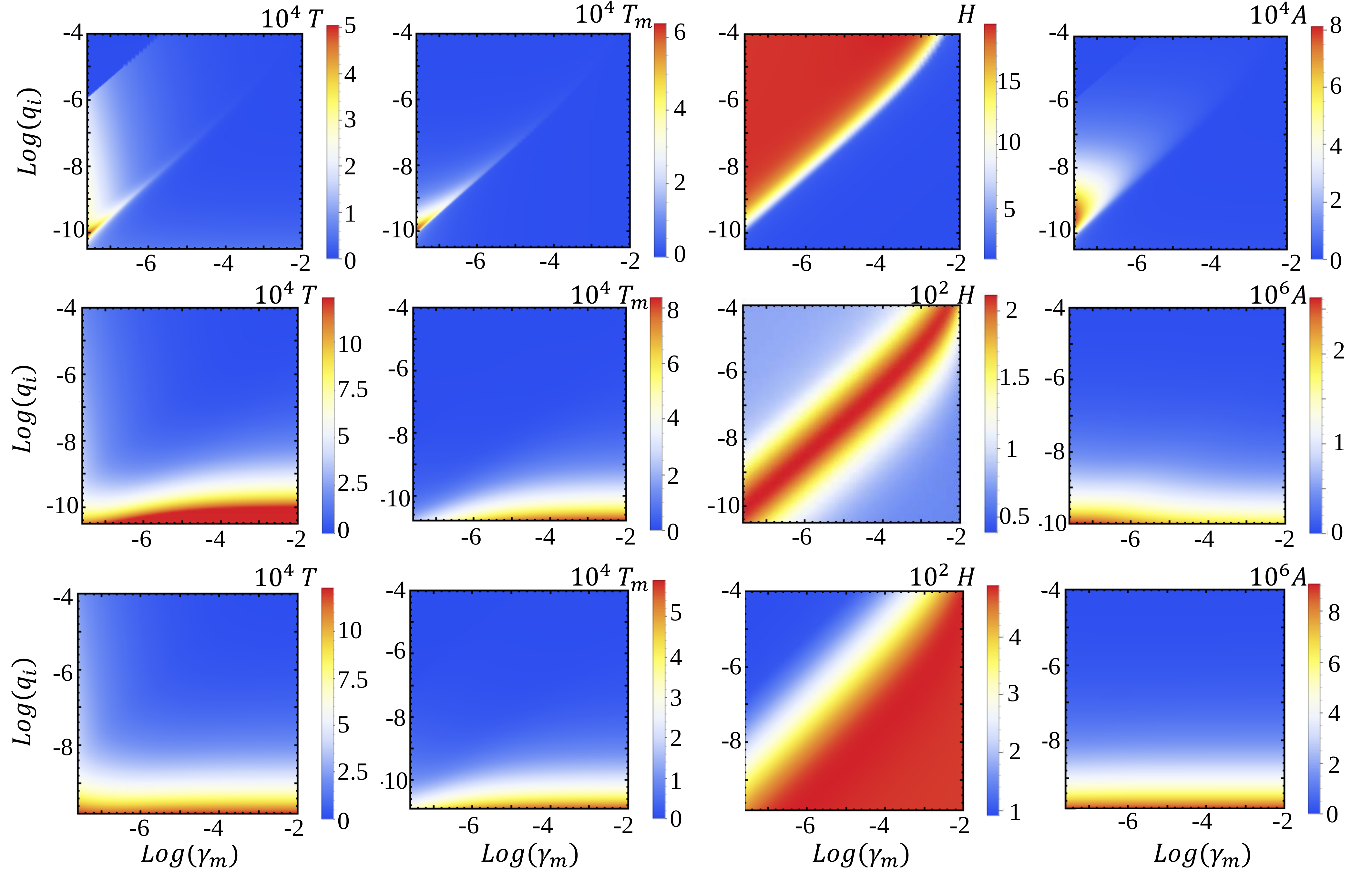}
\caption{(Color online) Duration, time-to-peak, severity and area associated to the infection profile $I(t)$ (from left to right: $T$, $T_{m}$, $H$ and $A$) as a function of $\gamma_{m}$ and $q_{i}$. For three values of $\lambda$ (from top to bottom: $0.022$, $0.15$, and $0.5$), $N=1000$ and $\gamma_{s}=0.1$.}
\label{CSC-Ext}
\end{figure*}
\subsubsection{Extensive features of infection profile $I(t)$}
We characterize the profile differences by looking at the \textit{extensive} features of $I(t)$. This becomes particularly useful when comparing with viral outbreaks in real complex systems since information about the microscopic parameters is typically unknown. We consider the duration of the outbreak which we call $T$; the peak of the infection, i.e. the maximum value of the number of infected $I(t)$ which we call $H$; the time to achieve this maximum, i.e. time-to-peak which we call $T_m$; and the area below the $I(t)$ curve which we call $A$. Figure \ref{CSC-Ext} shows the behavior of these extensive quantities, obtained by integrating numerically Eqs. \ref{SIReqsP2P}. Profile features are shown as a function of the mobility $\gamma_{m}$ and infection probability $q_{i}$ for three different values of the infection contact rate $\lambda$. The relationship between the duration, time to peak, and area becomes evident by showing the similar qualitative results for a given value of $\lambda$. Some key points emerge: (i) As $\lambda$ grows, the times (duration and time-to-peak) and area become independent of the mobility. (ii) As $q_{i}$ increases, the times and area become smaller. (iii) By increasing the parameter $\lambda$ the maximum height grows. (iv) The highest severity value $H$ shows linearity with $\gamma_{m}$ and $q_{i}$ (i.e. $q_i=e^3\gamma_m)$. (v) The regions in the $q_{i}$-$\gamma_{m}$ space where the maximum height is located, change from low mobility and high infection probability for small $\lambda$, to the region of low infection probability and high mobility for large $\lambda$. The transition between these two limits can be seen to occur around $\lambda=0.15$. (vi) For small values of $\lambda$, the times and area follow linearly their maximum value with $\gamma_{m}$ and $q_{i}$. Before the transition point at $\lambda=0.1$, this linearity is lost.
\begin{figure}
\centering
\includegraphics[width=0.95\linewidth]{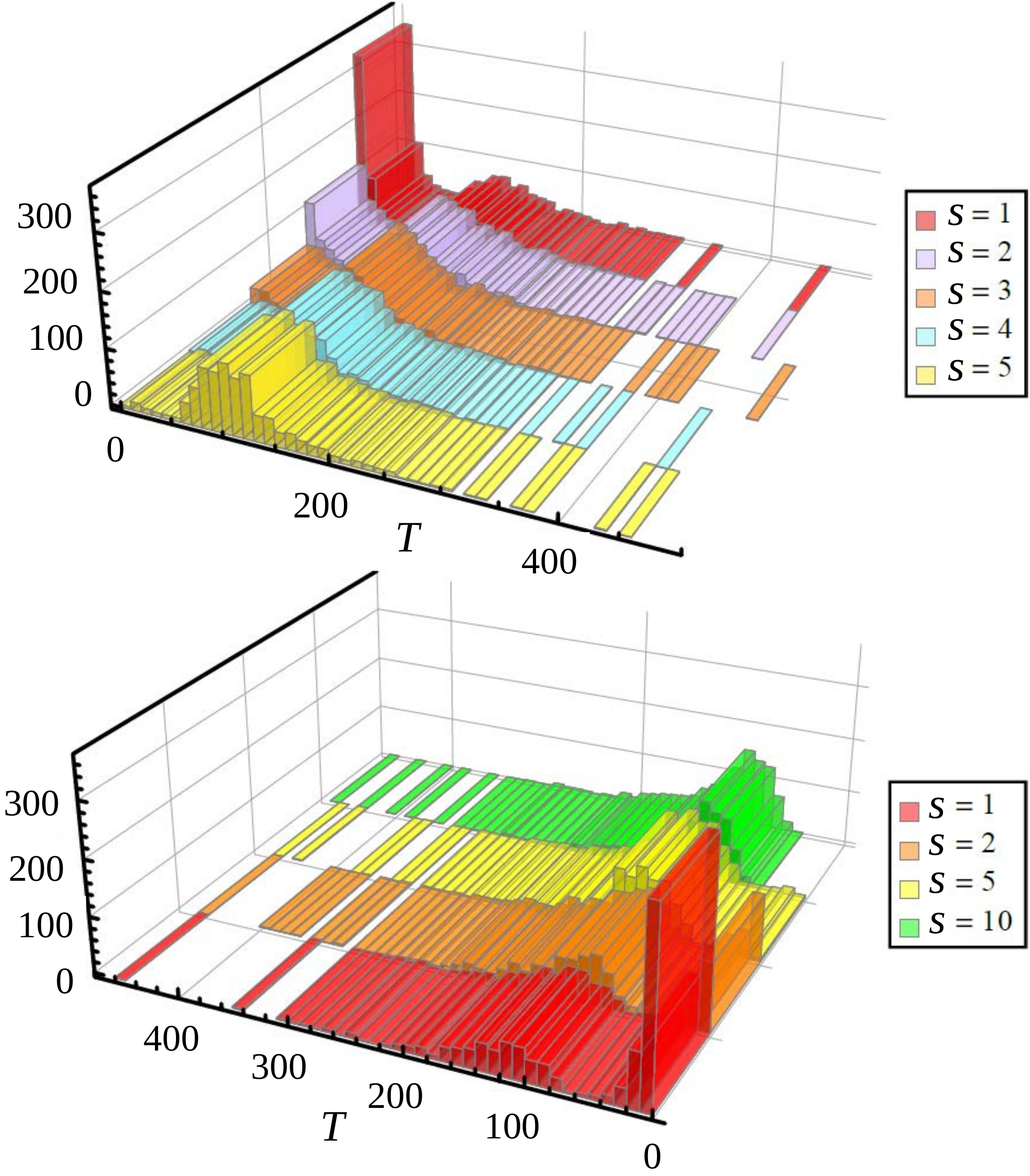}
\caption{(Color online) Distribution of duration of infection $T$, for different values of initial seed $s$. Parameters: $\lambda=0.022$, $q_{i}=0.002$, $\gamma_{m}=0.0018$ and $N=10^{3}$.}
\label{Dist-Blue}
\end{figure}

The initial conditions that we have so far considered, feature one infected individual in the space $G$. In real systems some infections are controlled or naturally dissipated before a large-scale spreading is reached. The numerical simulation can account for these types of situations: Figure \ref{Dist-Blue} illustrates the distribution of the infection's duration for $1000$ different realizations and different initial conditions which varies with the number of initially infected objects (seed $s$). Each run leads to a slightly different dynamics whose mean values are well captured by the dynamical equations. For small values of $s$, the probability of having a short outbreak ($T\approx0$) is very high in comparison with realizations for larger values of $s$. For this illustration, the recovery probability is selected to be approximately $50$ times greater than the infection probability. Hence, the distribution for small $s$ shows a large probability of a short infection, i.e. for most of the runs, the few infected agents recover faster than they can spread the infection. On the contrary as $s$ increases, the probability of short durations decreases and the distribution gets populated with a duration that is similar for all the values of $s$. Interestingly, the point where the distribution is maximum (after the short duration peak for small $s$) is only slightly shifted to shorter times as $s$ grows. It becomes more evident in the bottom of Fig. \ref{Dist-Blue} by looking at the difference between $s=5$ and $s=10$. The duration that more closely resembles the result from the differential equations, is around $T \approx 400$ (see blue curve in Fig. \ref{fig2}) and has a very low probability for all $s$ values. For this parameter choice, the simulation is far from the mean-field (i.e. differential equation) result.
\begin{figure}
\centering
\includegraphics[scale=0.56]{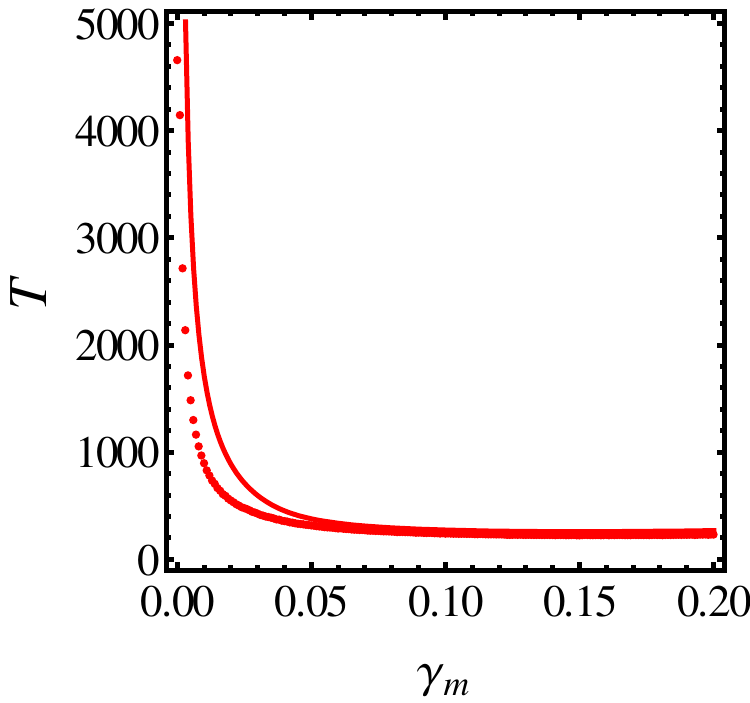}
\includegraphics[scale=0.56]{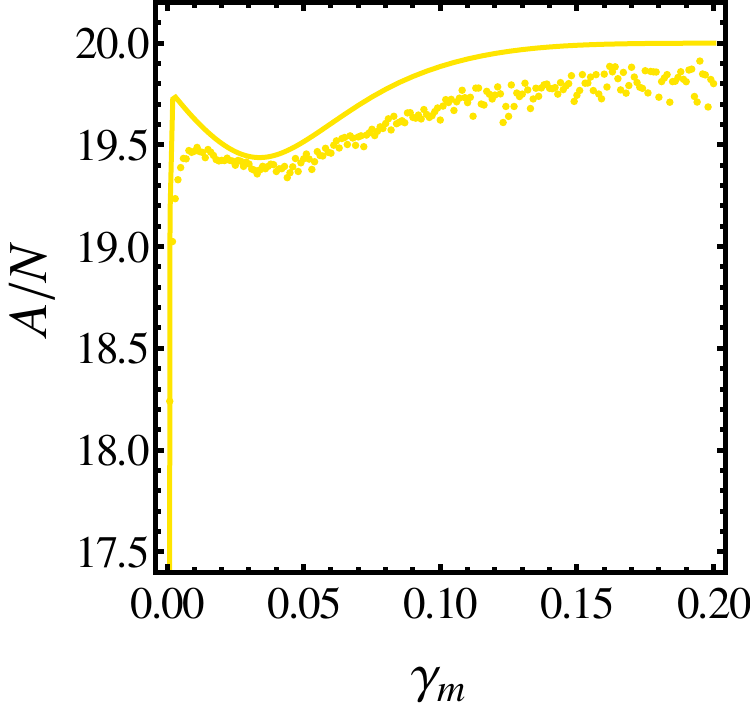}
\includegraphics[scale=0.56]{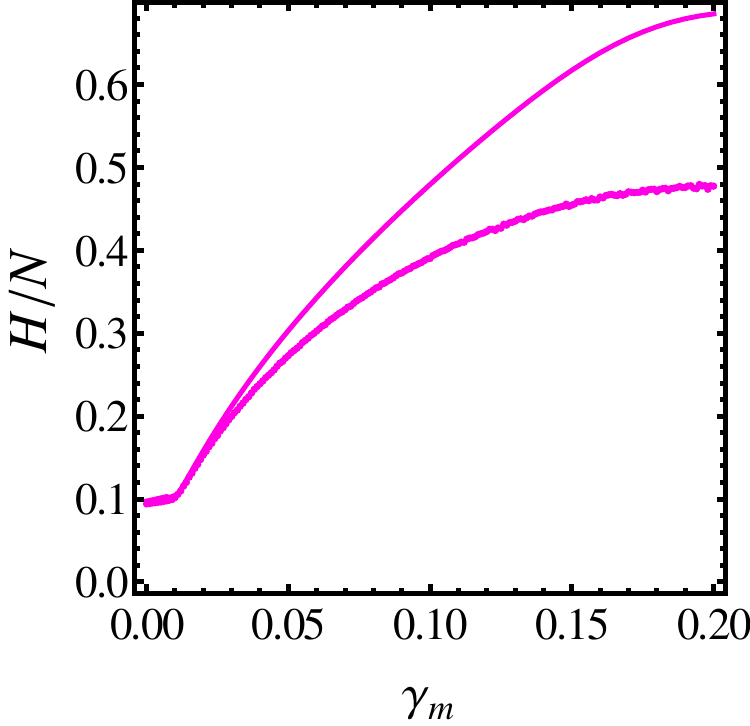}
\includegraphics[scale=0.56]{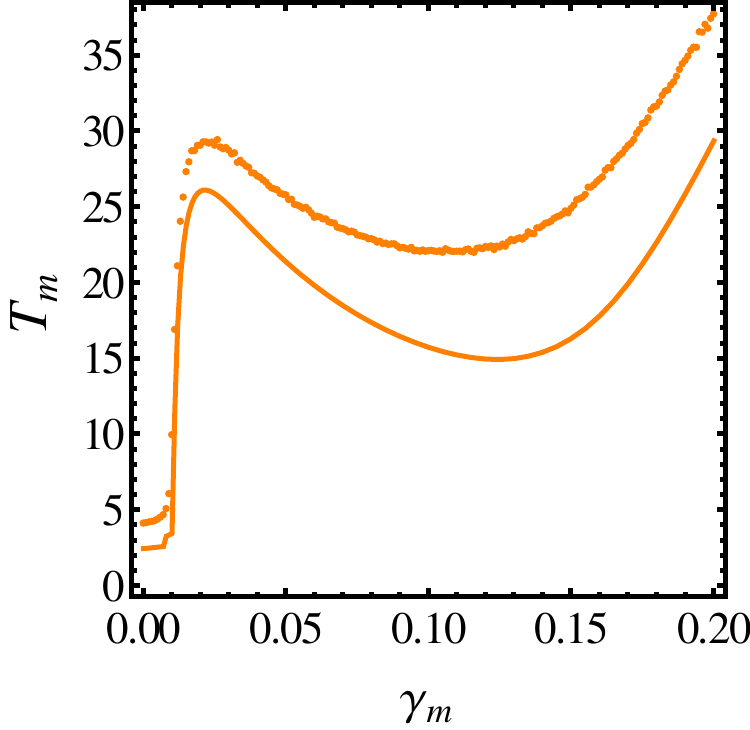}
\caption{(Color online) Duration (top left), severity (bottom left), area (top right) and time-to-peak (bottom right) as a function of mobility from numerical integration of differential equations (solid curve) and mean values of $10^4$ simulations (dotted curve). Parameters: $\lambda=0.1$, $q_{i}=0.005$, $\gamma_{s}=0.1$ and $N=10^{4}$.}
\label{Red-Contrast-Sim-DE}
\end{figure}

As an illustration, Fig. \ref{Red-Contrast-Sim-DE} depicts the result for the extensive quantities of the infection profile as a function of mobility, contrasting the results from the differential equations (solid curve) with the mean value from the simulations (dotted curve). The results show that for small values of mobility, the duration predicted by the differential equations is greater than the mean from simulation, in agreement with the previous finding. This statement is also valid for the area and time-to-peak, but it is false for the peak height. The latter displays, for small $\gamma_{m}$, a good agreement between the simulations and the equations. In contrast, as the mobility is increased, the statement is no longer accurate for the duration and maximum height. For instance, the simulation result grows with a smaller rate than the differential equation for the maximum height while the agreement between the results for duration grows as the mobility is increased.

\subsection{Comparison with real-world social contagion}

We note that the profiles in the top two insets of Fig. \ref{fig2} are commonly observed in association with the download popularity of YouTube clips reported in Refs. \cite{sornette1,riley}. The bottom inset is more characteristic of financial systems, and looks remarkably similar to the profile obtained for the revaluation of the Chinese Yuan currency reported in Ref. \cite{us2}. This same rumor circulated twice in the space of a few months, producing an almost identical profile. The currency pairs follow a similar dynamical pattern in each case, which suggests that the same underlying group dynamics developed, in line with our model \cite{us2}. 

\begin{figure}
\centering
\includegraphics[width=0.8\linewidth]{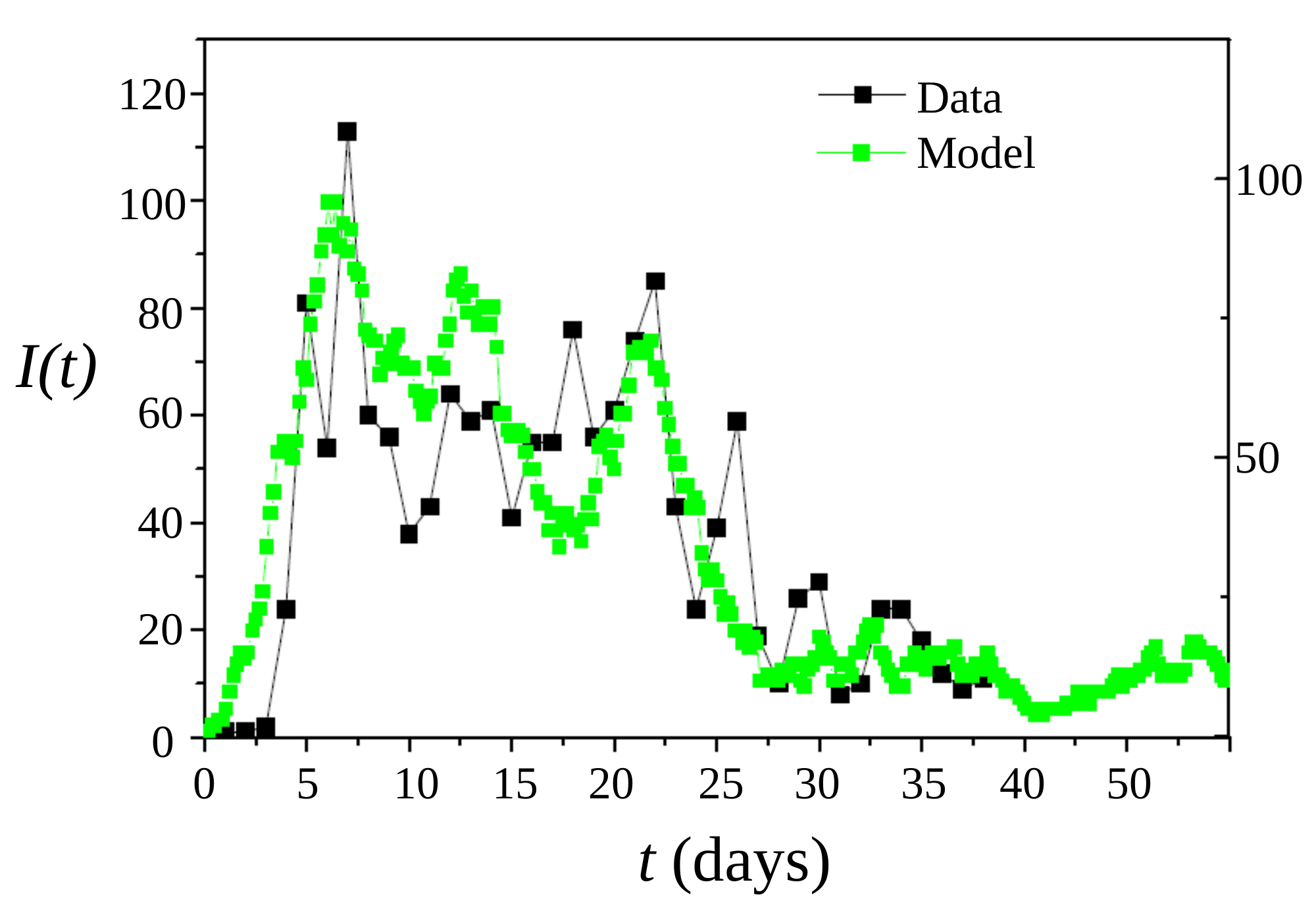}
\caption{(Color online) Infection profile from our model (green) and civil-unrest event profile in Libya during the Arab spring of 2011. We take $N = 4000$ agents and the timestep to be one day, starting on February 24, 2011. The other parameters are $p_j = 0.08$, $p_l = 0.72$ (for joining and leaving the group); $q_i = 0.1$, $q_r = 0.46$ (for infection and recovering processes in SIR).}
\label{libya}
\end{figure}

In the social domain, we also find similarities between the infection profiles produced by our dynamical model and those of civil-unrest events. The use of new technologies such as social media and mobile phones is arguably one of the main elements that helped large mobilizations such as the Arab Spring to generate continuous waves of civil unrest activity. Indeed the use of this technology during these protests doubled in some participant countries. The case of Libya is known to have relied on the use of cell phones, emails and YouTube videos to spread information about the current state of the protests and to coordinate new demonstrations \cite{stepoanova11}. Figure \ref{libya} compares the output of our model with the volume of civil unrest events in Libya during the 2011 Arab Spring, treating the events in a given day as a proxy for the number of infecteds $I(t)$. Given that our model accounts for the spreading dynamics within a population that is continuously renewed over time, individuals may join a popular place (e.g. enter a chatroom) and hence be susceptible to becoming infected (e.g. get influenced by a political idea). The dynamics of the number of infected individuals in our model is therefore likely to be illustrative of the on-street activity that then follows (e.g. demonstration events). While we are well aware of the limitations in making such a connection, it is nevertheless a reasonable first approximation. As can be seen from the results in Fig. \ref{libya}, the model mechanisms such as mobility and spreading result in revivals on the infected community that matches reasonably well the actual on-street event data. We stress that while these profiles for the model and actual on-street events shown in Fig. \ref{libya} are similar to each other, each is very different from that predicted by a well-mixed standard SIR model which is characterized by exponential decays and no revivals.

\begin{figure}
\includegraphics[width=0.7\linewidth]{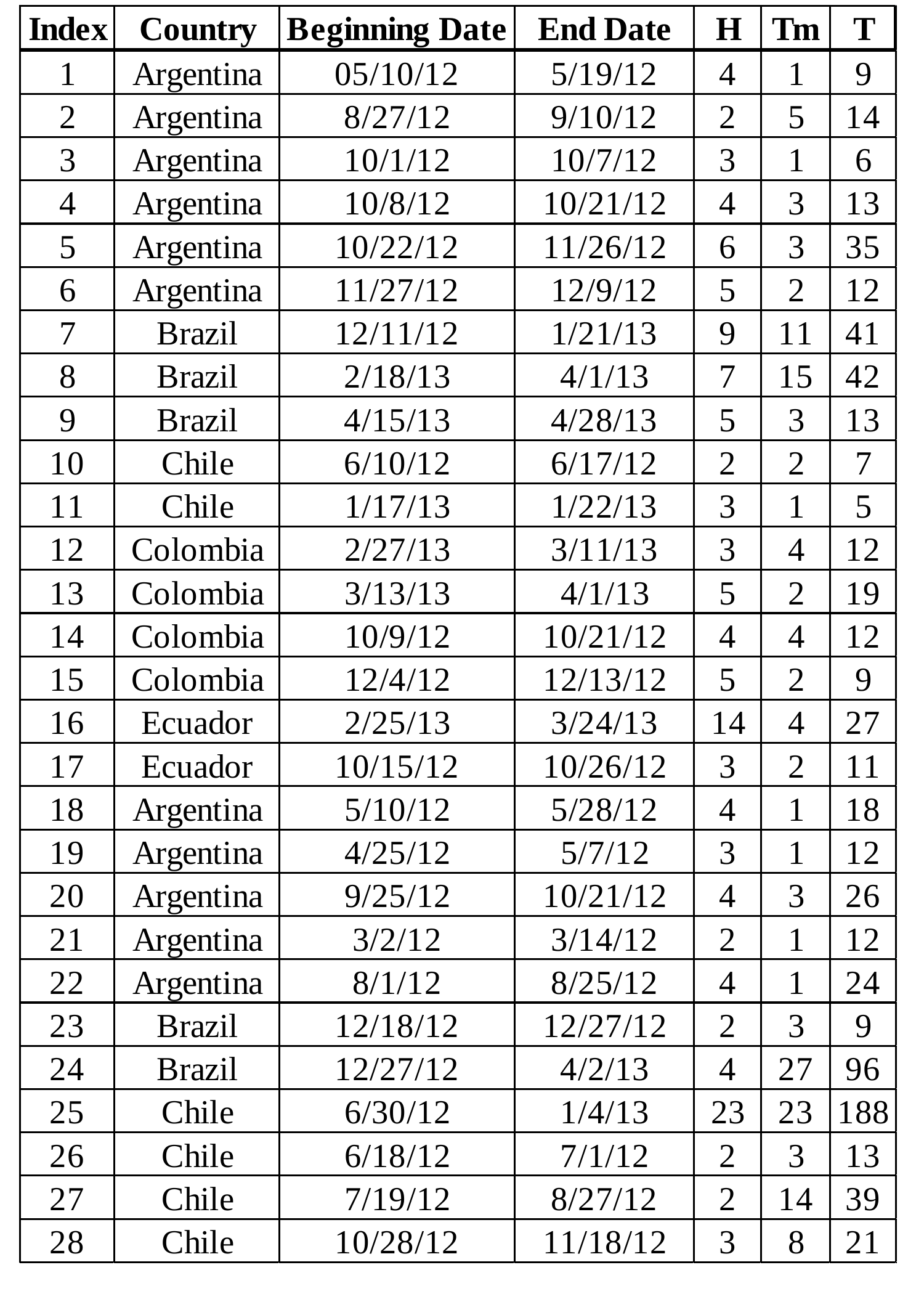}
\includegraphics[width=0.7\linewidth]{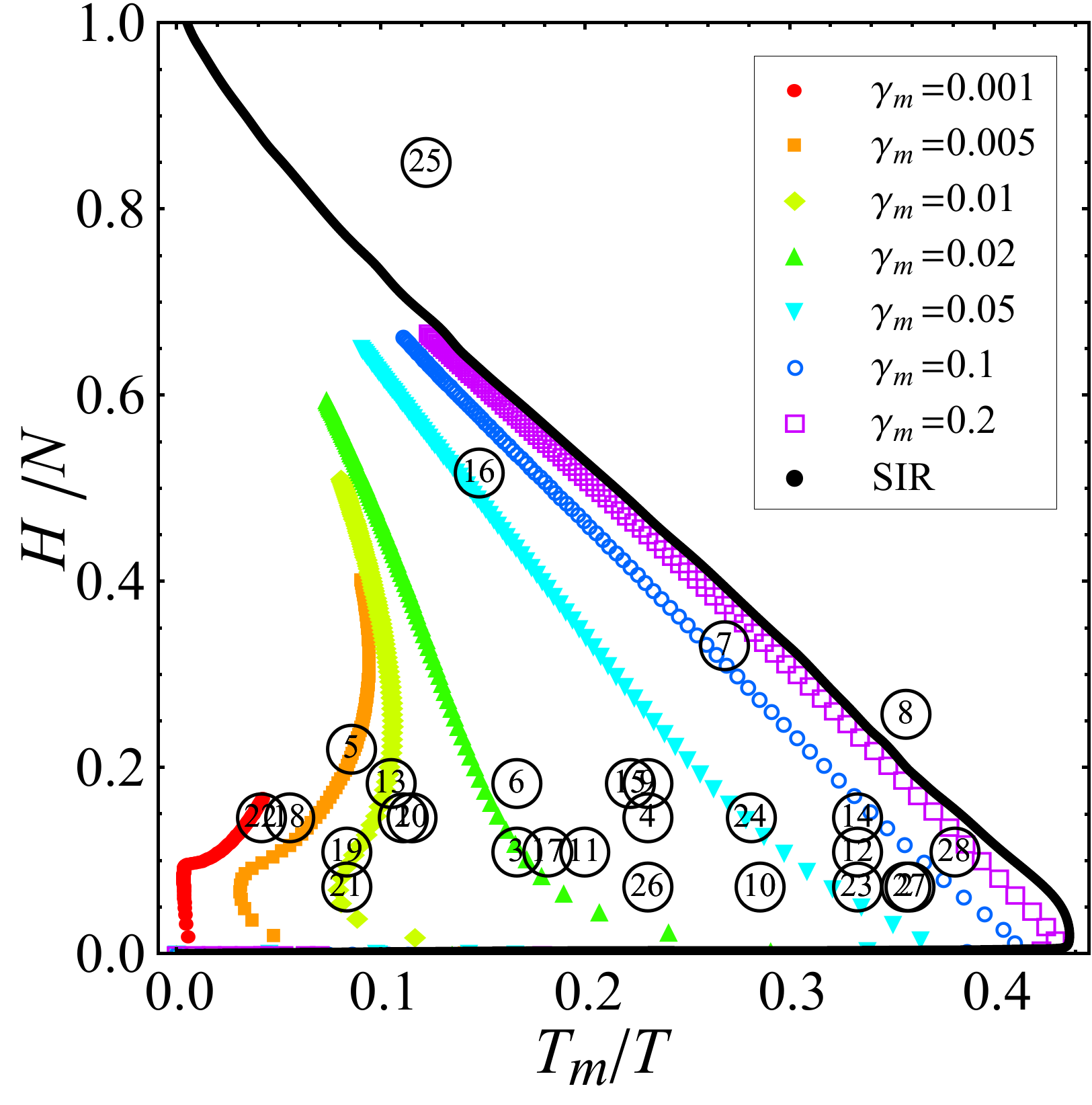}
\caption{(Color online) (Top) Bursts of civil unrest. Columns show burst index, start time, end time, and bursts' features in days ($T$ and $T_m$) and number of events ($H$). (Bottom) Outbreak profile descriptors $H/N$ and $T_m/T$ compared to empirical data of on-street civil unrest (numbered circles) \cite{IARPA} from the numbered top list. Theoretical lines obtained by integrating the coupled differential equations for different values of mobility $\gamma_m$. Thick black line shows result for standard (i.e. well-mixed) SIR model. $N=1000$, $q_i = 0.002$ throughout. Each trajectory starts near origin for $\lambda\equiv q_i/q_r=10^{-3}$ and grows until $\lambda=1$ in steps of $\delta\lambda = 10^{-3}$. To estimate $N$ from the real data, we chose the $97\%$ quantile of $H$ (i.e. a z-score of $3$) from a larger sample and obtained a quantile of $27$.}
\label{CSC-Ext2}
\end{figure}

Figure \ref{CSC-Ext2} goes further by comparing the extensive features of the model's infection profile to that obtained from empirical data of on-street civil protest events in Latin America (numbered circles). Again we are taking the number of infected $I(t)$ at a given timestep as a proxy for the number of people incited to protest, and the space $G$ as some physical or even online space (e.g. city center or chatroom) where individuals become sufficiently motivated to protest. While we are not suggesting it provides a unique or definitive explanation of these phenomena, the model (thin colored lines) does  capture the wide variability of outbreak profiles in a way that a standard SIR model cannot (thick black line). 
The on-street civil unrest data (numbered circles) come from a unique multi-year, national research project involving exhaustive event analysis by subject matter experts (SMEs) across an entire continent (see Refs. \cite{unrest,IARPA}). The start and end of each burst is identified using the analysis of Ref. \cite{Karsai2} and cross-checked manually. The key to extract features from the sequence of events, is to construct the infection profile(s). First, we segment a long sequence of civil unrest events by a pre-specified threshold $d$. That is, if the interval between two consecutive events is not larger than $d$, the latter event is in the same segment as the previous event. Second, the curve of infection is built based on one segment of events, by making the reciprocal of the intervals between events as the $y$ values and the time step as the $x$ value. Then we can extract the features of that infection curve, forming one numbered circle in Fig. \ref{CSC-Ext}(bottom). 

\subsection{Broadcast SIR mechanism}
In the broadcast-type infection model, the space $G$ has a constant infection rate $q_i$ for infecting the susceptible who happen to be in $G$ at that timestep -- which is akin to having contaminated surfaces in a hospital, school or airport. All the infecteds have a recovery rate $q_r$ to recover and become immune. These dynamics are governed by the equations:
\begin{eqnarray}
\frac{dS_g}{dt}&=&-q_i S_g -p_l(S_g - q_i S_g)+p_j (S-S_g),\nonumber\\
\frac{dI_g}{dt}&=&q_i S_g-q_r I_g - p_l(I_g+q_i S_g-q_r I_g)\nonumber\\
&&+(1-q_r)p_j(I-I_g),\nonumber\\
\frac{dR_g}{dt}&=&q_rI_g-p_l(R_g+q_r I_g)+p_j((R-R_g)\nonumber\\
&&+q_r(I-I_g)),\nonumber\\
\frac{dS}{dt}&=&-q_iS_g,\nonumber\\
\frac{dI}{dt}&=&q_iS_g-q_rI,\nonumber\\
\frac{dR}{dt}&=&q_rI.
 \label{SIReqsBC}
\end{eqnarray}
\begin{figure}
\centering
\includegraphics[width=0.9\linewidth]{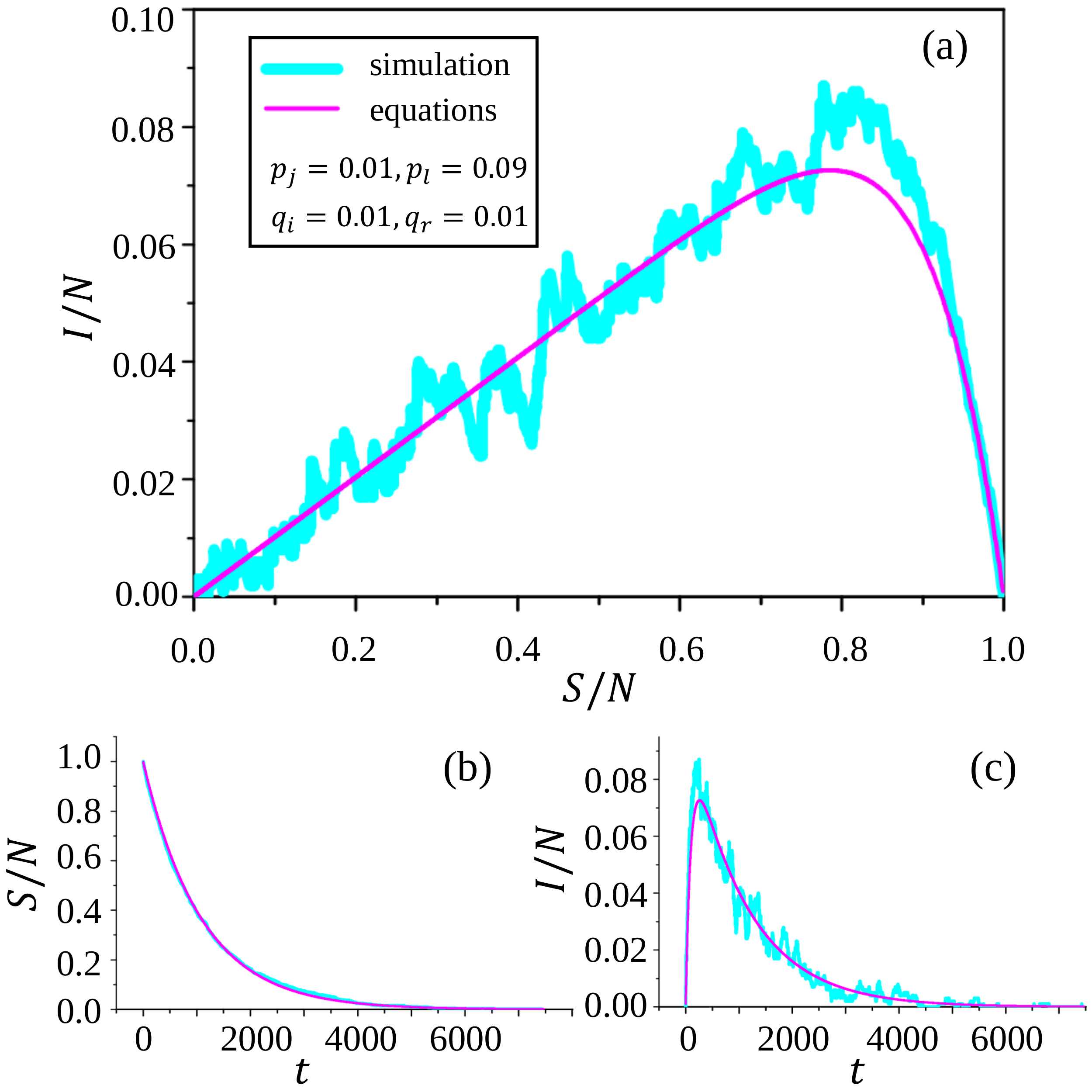}
\caption{(Color online) (a) Trajectories of evolution of the system in the $S-R$ space. Initially the system starts from the right-hand lower corner. (b) The evolution of $S(t)$ in time, and (c) the evolution of
$I(t)$ in time. The rougher and smoother curves are obtained by simulation and by integrating the set of equations (Eq. (\ref{SIReqsBC})) respectively. The systems have the same mean number of agents in $G$ given by $N \gamma_s$ where $\gamma_s = 0.1$. Other parameters are: $\lambda = 1.0$, $q_i = 0.01$, $p_j = 0.01$ and $p_l = 0.09$.}
\label{sir_bc}
\end{figure}
An explicit formula for $S(t)$ can be obtained by solving the first and fourth equations in Eq. (\ref{SIReqsBC}). The result is
\begin{eqnarray}
S(t)&=&C_1\exp\left\{\frac{1}{2}(-A-\sqrt{A^{2}-4B})t\right\}\nonumber\\&&+C_{2}\exp\left\{\frac{1}{2}(-A+\sqrt{A^{2}-4B})t\right\}
\label{SolS_BC}
\end{eqnarray}
where $A = ((p_j + p_l ) + (1-p_l )q_i )$, $B = p_j q_i $, and $C_1$ and $C_2$ are determined by
the initial conditions. Equation (\ref{SolS_BC}) shows that the decrease of susceptibles is not related to the
number of infecteds, i.e. $q_r$ is irrelevant. This is in contrast with the person-to-person
case, where the results depend significantly on $q_r$. 

\begin{figure}
\centering
\includegraphics[width=0.95\linewidth]{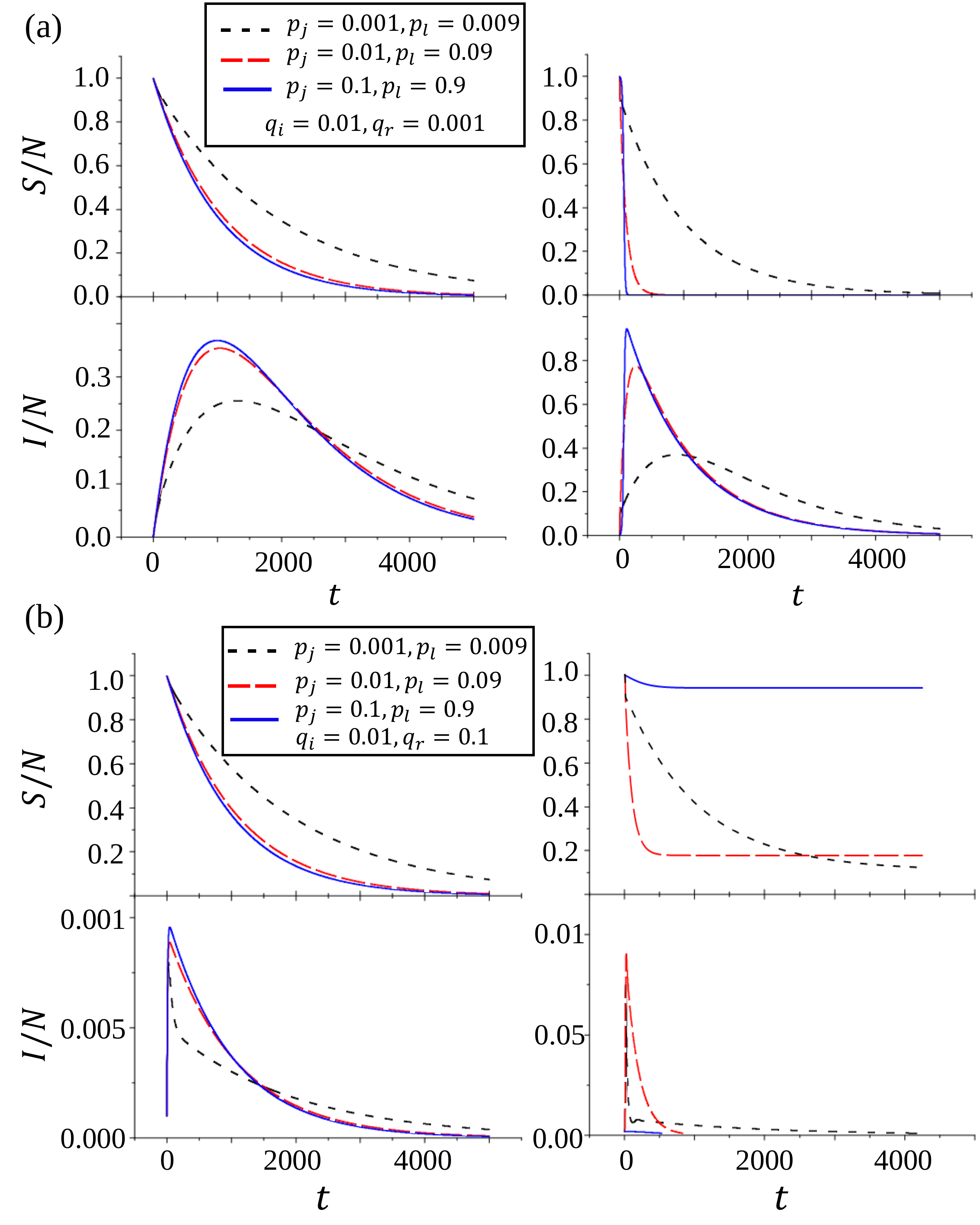}
\caption{(Color online) Comparison between the broadcast (left column) and person-to-person (right column) cases. (a) $q_i > q_r$; (b) $q_i < q_r$. Results obtained by solving Eq. (\ref{SIReqsP2P}) for the person-to-person case and Eq. (\ref{SIReqsBC}) for the broadcast case. Parameters are shown in the figure.}
\label{num-si}
\end{figure}

\begin{figure}
\centering
\includegraphics[width=0.95\linewidth]{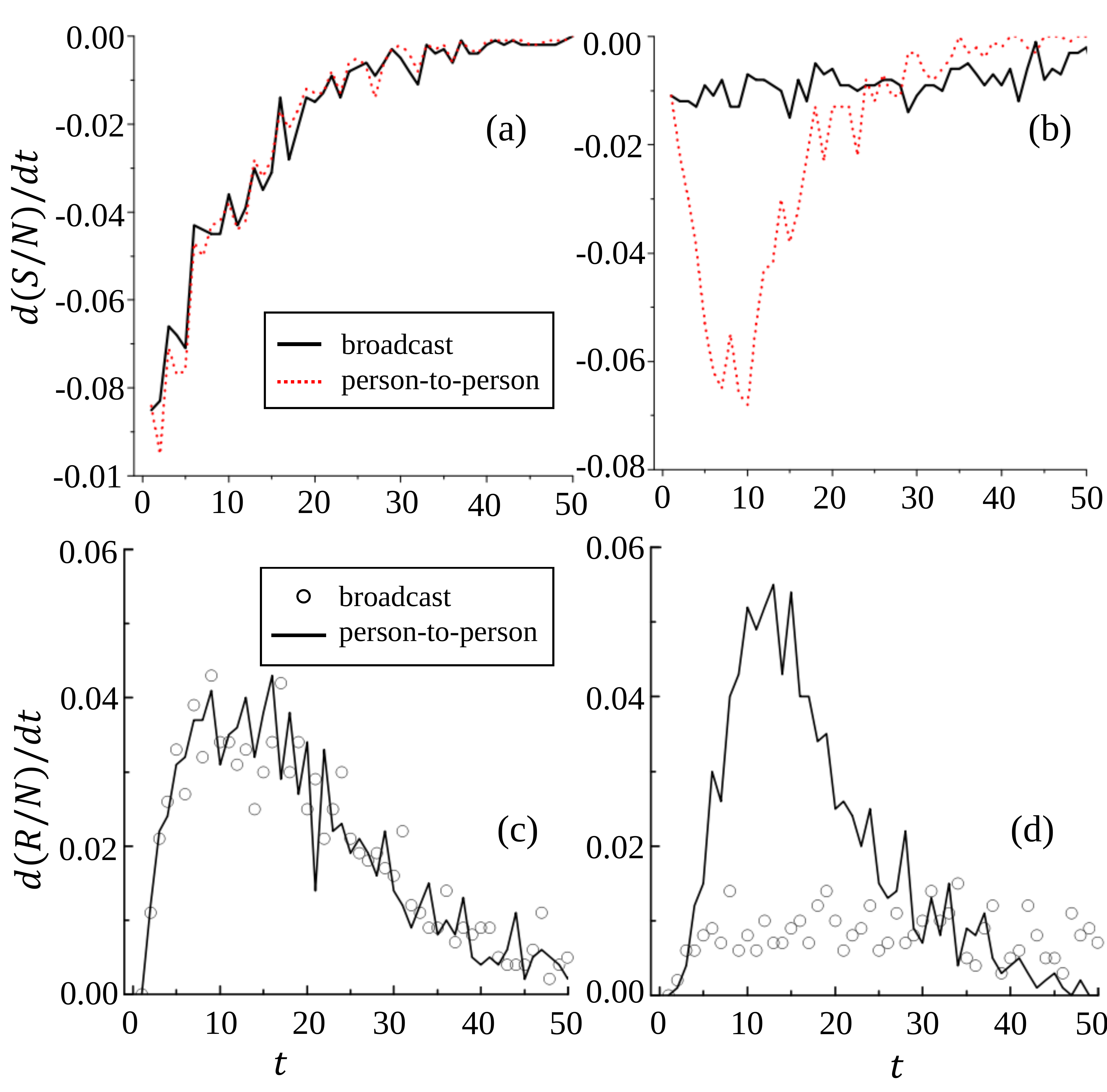}
\caption{(Color online) (a-b) Derivative $d(S(t)/N)/dt$ and (c-d) derivative $d(R(t)/N)/dt$ as a function of time for the broadcast and person-to-person infection mechanisms obtained by numerical simulations. The parameters are: $p_j = 0.1$, $p_l = 0.9$, (a) and (c) $q_i = 0.9$, $q_r = 0.1$; and, (b) and (d) $q_i = 0.1$, $q_r = 0.2$.}
\label{sim-dsdr}
\end{figure}

Figure \ref{sir_bc} shows the results of this dynamical process using the equations and direct simulation. The two sets of results are basically consistent with each other. Figure \ref{num-si} compares results for the broadcast-type infection and person-to-person infection, by solving Eq. (\ref{SIReqsP2P}) and Eq. (\ref{SIReqsBC}). Comparing the behavior in the two cases for $q_i > q_r$ (Fig. \ref{num-si}(a)), we note that there is a difference in behavior in both $S(t)$ and $I(t)$ at short times. This difference is more apparent for the case of $q_i < q_r$ , as shown in Fig. \ref{num-si}(b).
In order to analyze further the two infection mechanisms, we look at the quantities $(1/N)(dS/dt)$ and $(1/N)(dR/dt)$. Figure \ref{num-si} shows the results obtained by solving Eq. \ref{SIReqsP2P} (for person-to-person infection) and Eq. \ref{SIReqsBC} (for broadcast infection). Figure \ref{sim-dsdr} shows the results for the person-to-person and broadcast mechanism obtained by simulations. For the cases of Fig. \ref{sim-dsdr}(a) and Fig. \ref{sim-dsdr}(c) where the infection probability is high and the recovery probability is low ($q_i = 0.9$ and $q_r = 0.1$), the behaviors for the two infection mechanisms are similar. This is because the parameters correspond to the situation where a susceptible getting into $G$ will almost certainly become infected, regardless of the infection mechanism (i.e. very high infection probability in comparison with recovery). However, for the cases in Fig. \ref{sim-dsdr}(b) and Fig. \ref{sim-dsdr}(d) where the infection probability is low and the recovery probability is high ($q_i = 0.1$ and $q_r = 0.2$), the two mechanisms give different behaviors with the broadcast mechanism showing a less variable $dR/dt$.

\begin{figure*}
\centering
\includegraphics[width=0.9\linewidth]{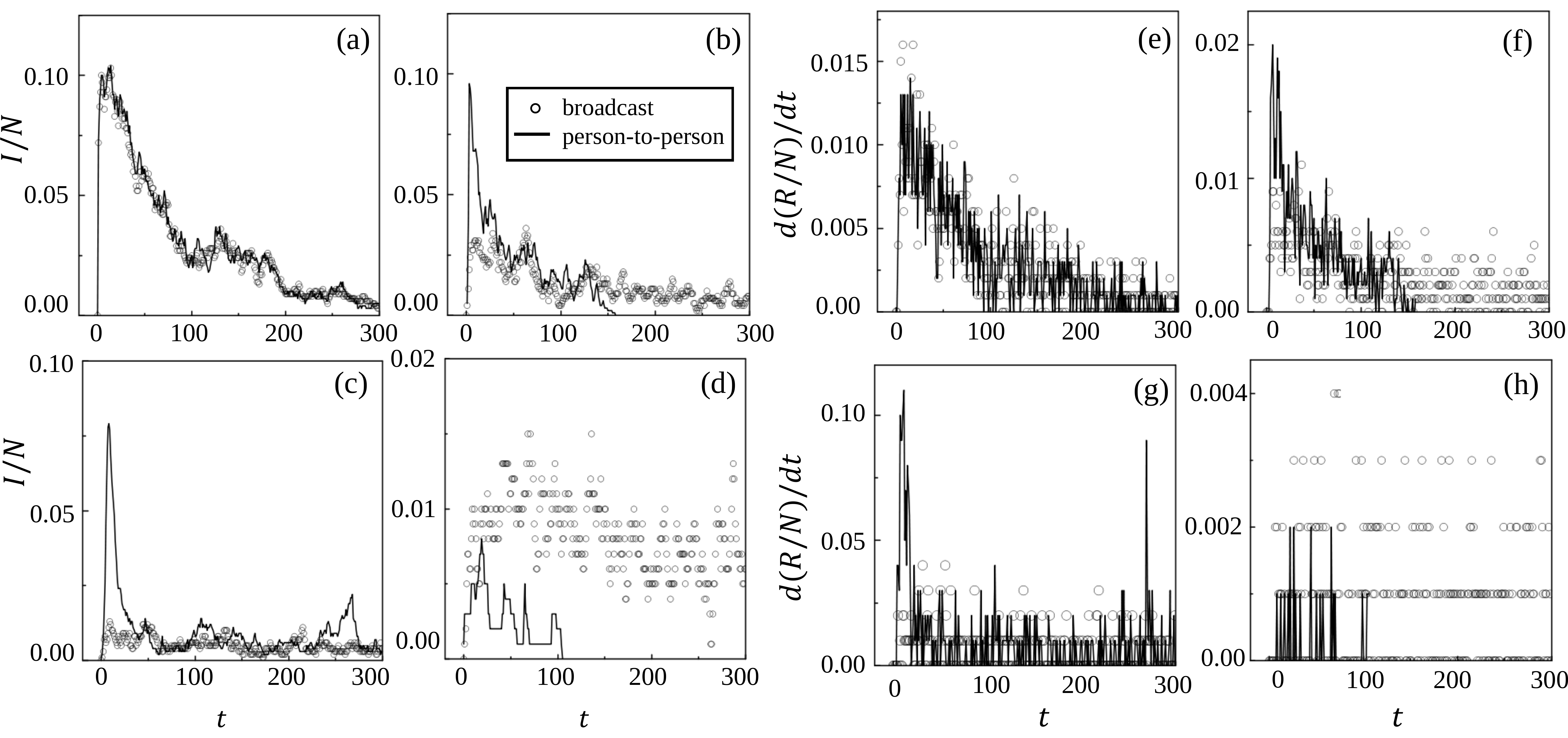}
\caption{The fraction of infected individuals $I(t)/N$ and the derivative $d(R(t)/N)/dt$ as a function of time for the broadcast and person-to-person mechanism. The parameters are: (a) and (e) $\gamma_s = 0.1$, $\gamma_m = 0.018$, $q_i = 0.9$, $q_r = 0.1$, (b) and (f) $\gamma_s = 0.1$, $\gamma_m = 0.018$, $q_i = 0.1$, $q_r = 0.2$, (c) and (g) $\gamma_s = 0.1$, $\gamma_m = 0.0018$, $q_i = 0.01$, $q_r = 0.1$, (d) and (h) $\gamma_s = 0.1$, $\gamma_m = 0.18$, $q_i = 0.01$, $q_r = 0.1$. (a-b) and (e-f) correspond to fixed average group size and mobility, but different infection and recovery probabilities. (c-d) and (g-h) correspond to fixed infection and recovery probabilities, but different mobilities.}
\label{sim-Idr}
\end{figure*}

Figures \ref{sim-dsdr} and \ref{sim-Idr} compare simulation results of $I(t)/N$ and $d(R(t)/N)/dt$ for the two infection mechanisms, for several different sets of model parameters. Figure \ref{sim-Idr}(a) and (b) (Fig. \ref{sim-dsdr}(c) and (d)) correspond to cases with fixed average occupation of $G$ and mobility but with different infection and recovery probabilities. When the infection probability is high and the recovery probability is low, the two mechanisms give similar results (Fig. \ref{sim-Idr}(a) and Fig. \ref{sim-dsdr}(c)). When the infection probability is higher than the recovery probability ($\lambda = q_i /q_r = 2.0$) as in Fig. \ref{sim-Idr}(b) and Fig. \ref{sim-dsdr}(d), the number of infecteds increases faster in the early stage for the person-to-person mechanism. Figures \ref{sim-Idr}(c-d) and (g-h) correspond to cases with fixed parameters for the viral process (fixed $q_i$ and $q_r$ ), but different parameters in terms of joining and leaving the space $G$. In Fig. \ref{sim-Idr} (c) and (g), the mobility is low ($\gamma_m = 0.0018$). 

Although the infection probability is small, the person-to-person infection mechanism still shows a strong epidemic at short times and then an oscillatory behavior. For the same infection probability, the broadcast mechanism shows a weak epidemic. Thus, the low mobility enhances the epidemic even though the infection probability is small, by retaining and thus increasing the number of infecteds within $G$ who can then further infect susceptibles. This infection reinforcement mechanism is missing from the broadcast case. In Figs. \ref{sim-Idr}(d) and (h), the mobility is high ($\gamma_m = 0.18$ with the same $q_i$ and $q_r$ as in (c)) and the person-to-person mechanism does not cause an epidemic, however the broadcast mechanism does lead to an epidemic. Whether an infection is spread through personal contact or through contact with the physical space itself (e.g. contaminated surfaces) is therefore crucial in dictating the infection profile $I(t)$ to be expected.


\section{SIS process}
\subsection{Person-to-person SIS mechanism}
\begin{figure}[tbp]
\begin{center}
\includegraphics[width=0.95\linewidth]{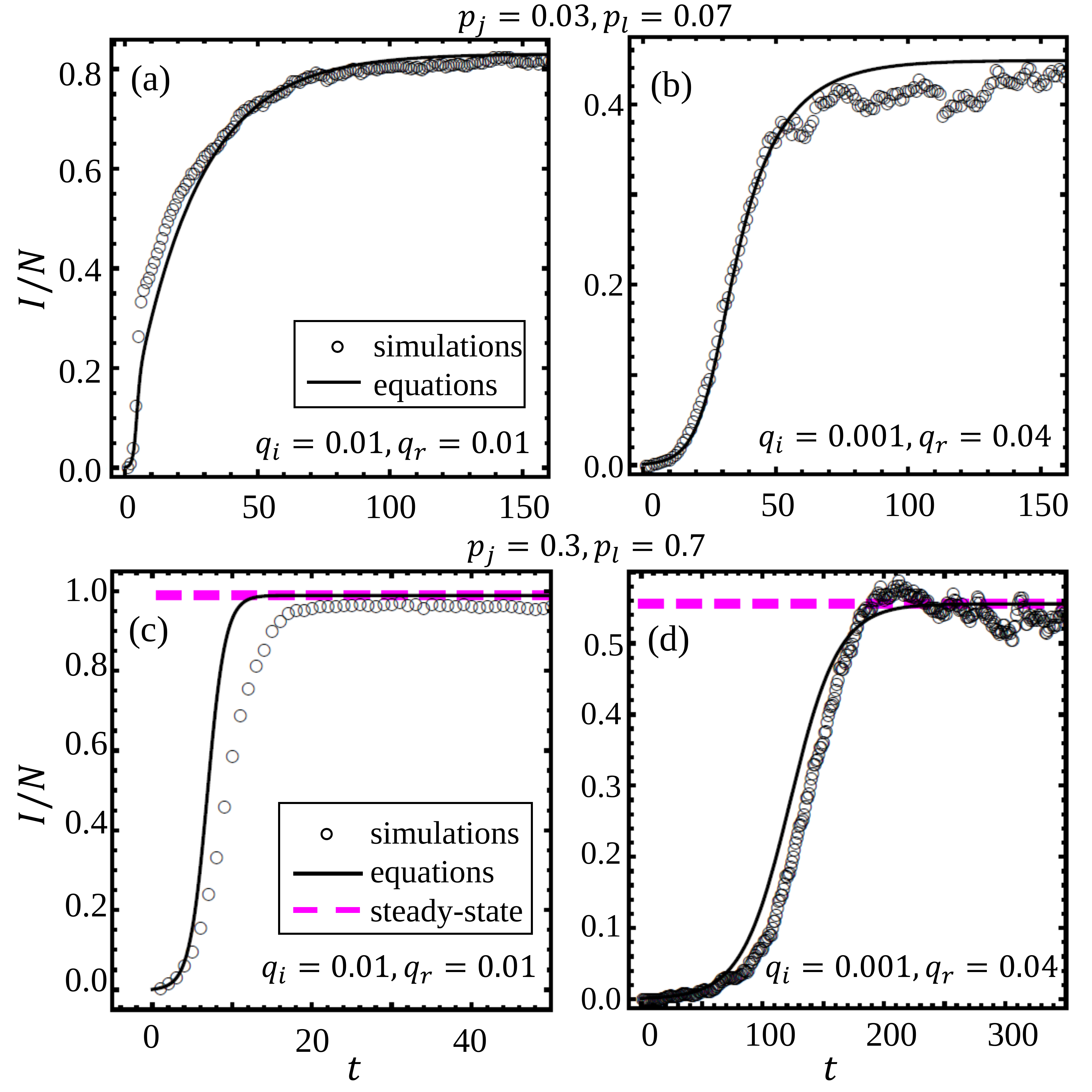}
\end{center}
\caption{(Color online) Fraction $I/N$ as a function of time $t$, for SIS model with person-to-person infection mechanism. Several sets of parameters are selected to illustrate the features.}
\label{fig:ptp}
\end{figure}

We now study the same mobility model involving $G$, but now using an SIS (Susceptible-Infected-Susceptible) viral process. As before, infected individuals can only infect others when they are present in the space $G$, and each infected in $G$ has a probability $q_i$ (per unit step) to infect a susceptible in $G$. All the infected individuals in the system (inside and outside $G$) have a probability $q_r$ to recover and become susceptible again. At the beginning of the process, we randomly select an agent in $G$ to be infected. The viral dynamics of the system are governed by the following equations in the mean-field limit:

\begin{eqnarray}
 \frac{{dS_g }}{{dt}} &=&  - q_i S_g I_g  + q_r I_g  - p_l (S_g  - q_i S_g I_g  + q_r I_g )\nonumber\\
&& + p_j (S - S_g  + q_r (I - I_g )) \nonumber \\
 \frac{{dI_g }}{{dt}} &=& q_i S_g I_g  - q_r I_g  - p_l (I_g  + q_i S_g I_g  - q_r I_g )\nonumber \\ 
&&+ p_j (1 - q_r )(I - I_g ) \nonumber \\
 \frac{{dS}}{{dt}} &=&  - q_i S_g I_g  + q_r I \nonumber \\
 \frac{{dI}}{{dt}} &=& q_i S_g I_g  - q_r I
 \label{SISeqs}
\end{eqnarray}
For the sake of brevity, we focus on a few choice sets of parameters $(q_i,q_r)$ and $(p_j,p_l)$, for both the simulations ($N=1000$ agents) and for the numerical integration of Eq. (\ref{SISeqs}).  Results are shown in Fig. \ref{fig:ptp}. The numerical integration results agree well with the simulation results showing a monotonic increase in the fraction of infected individuals for recovery probabilities equal to and larger than the infection probability.
To fully describe the dynamical process, we then add the master equation concerning the dynamics in $G$:
\begin{equation}
 \frac{d N_g}{dt}= -p_l(S_g+I_g)+p_j(S-S_g+I-I_g).
 \label{GRPeq}
\end{equation}

The steady state (e.g. $S(\infty)$) may be found by setting the right-hand side of the equations (Eqs. (\ref{SISeqs})-(\ref{GRPeq})) to zero. We then obtain  
\begin{equation}
S_g({\infty})= S(\infty)\gamma_s,
 \label{Sg}
\end{equation}
\begin{equation}
I_g({\infty})= I(\infty)\gamma_s,
 \label{Ig}
\end{equation}
\begin{eqnarray}
S({\infty}) &=& N + \frac{1}{{2q_r A}}N\gamma _s ( - B + \sqrt {B^2
- C} )\nonumber \\ &&+ \frac{1}{{4q_i q_r A^2 }}( - B + \sqrt {B^2  - C} )^2,
 \label{S}
\end{eqnarray}
where
\begin{equation}
A = p_j + q_r (1 -(p_j + p_l )),
\end{equation}
\begin{equation}
B = N\gamma _s q_i A + q_r A + p_l q_r,
\end{equation}
\begin{equation}
C = 4Nq_i q_r A(p_j + \gamma _s q_r (1 - (p_j + p_l ))).
\end{equation}
So far, this steady-state solution is general.  For the special case scenario discussed earlier in which $p_j+p_l=1$, we obtain $\gamma_{s} = p_{j}$ and $\gamma_{m} = 2p_{j}(1-p_{j}) = 2 \gamma_{s}(1-\gamma_{s})$. In this case:
\begin{equation}
S(\infty)=\frac{q_r}{q_i \gamma_s^2}.
\end{equation}
Though the total number $N$ disappears from the limit $S(\infty)$, $N$ still enters as a bound. More explicitly, the above expression should be written as 
\begin{equation} 
S(\infty)=Min[\frac{q_r}{q_i \gamma_s^2},N].
\end{equation}
The corresponding value of $I(\infty)/N$ in the case of
$p_j+p_l=1$ is
\begin{equation}
\frac{I(\infty)}{N}=1-Min[\frac{q_r}{Nq_i \gamma_s^2},1].
\end{equation}
As an example of accuracy, Fig.\ref{fig:ptp}(c-d) compares the steady state $I(\infty)/N$ with the outcome from the differential equation as well as simulations for two sets of parameters that fulfill the condition $p_j+p_l=1$. In fact, under this special condition, the master equation for $S(t)$ becomes
\begin{equation}
\frac{{dS}}{{dt}} =  - q_i \gamma _s^2 SI + q_r I,
\label{eff-well-mixed}
\end{equation}
and is easy to solve for $S(t)$.  Equation (\ref{eff-well-mixed}) corresponds to an effective (well-mixed) SIS system. Substituting $I=N-S$ into the above equation, we obtain two solutions. One solution is (for decaying $I$)
\begin{equation}
S(t) = \frac{{q_r e^{Nq_i \gamma _s^2 t + q_r C}  - Ne^{q_r t +
Nq_i \gamma _s^2 C} }}{{q_i \gamma _s^2 e^{Nq_i \gamma _s^2 t +
q_r C}  - e^{q_r t + Nq_i \gamma _s^2 C} }} \;.
\end{equation}
If the initial condition is $S(0)=N_0$, the constant $C$ is given by
\begin{equation}
C = \frac{{\ln (q_r  - N_0 q_i \gamma _s^2 )}}{{Nq_i \gamma _s^2 - q_r }}.
\end{equation}
The other solution is (for increasing $I$)
\begin{equation}
S(t) = \frac{{q_r e^{Nq_i \gamma _s^2 (t + C)}  + Ne^{ - q_r (t - C)} }}{{q_i \gamma _s^2 e^{Nq_i \gamma _s^2 (t + C)}  + e^{q_r (t + C)} }} \;.
\end{equation}
If the initial condition is $S(0)=N_0$, the constant $C$ is given by
\begin{equation}
C = \frac{{\ln (N_0 q_i \gamma _s^2  - q_r )}}{{q_r  - Nq_i \gamma_s^2 }}.
\end{equation}
\subsection{Broadcast SIS mechanism}
\begin{figure}[tbp]
\begin{center}
\includegraphics[width=0.95\linewidth]{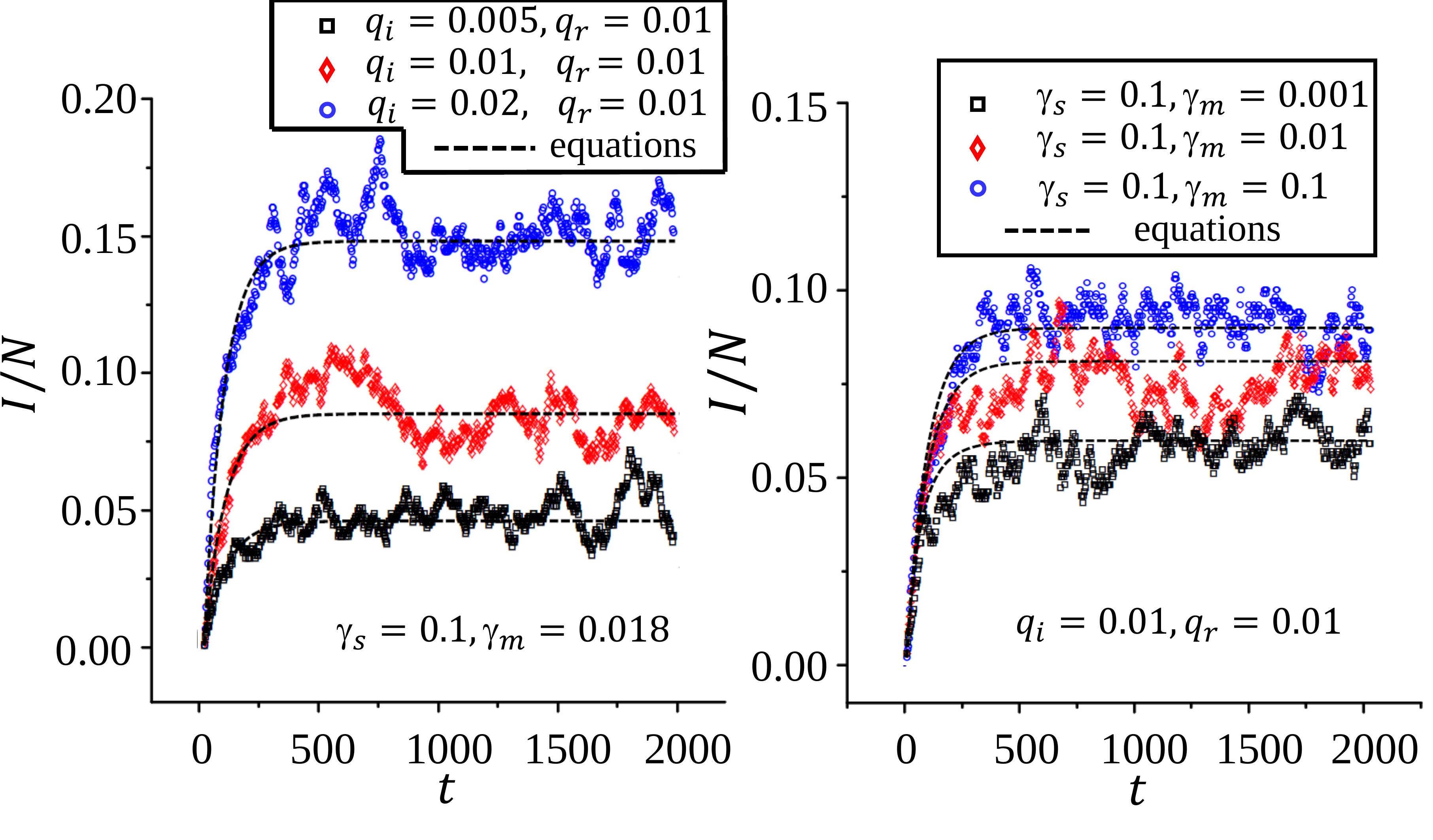}
\end{center}
\caption{(Color online) Fraction $I/N$ as a function of time $t$, for SIS model with a broadcast-type infection mechanism.  Several sets of parameters are selected to illustrate the features.}\label{fig:bd01}
\end{figure}

\begin{figure}[tbp]
\begin{center}
\includegraphics[width=0.95\linewidth]{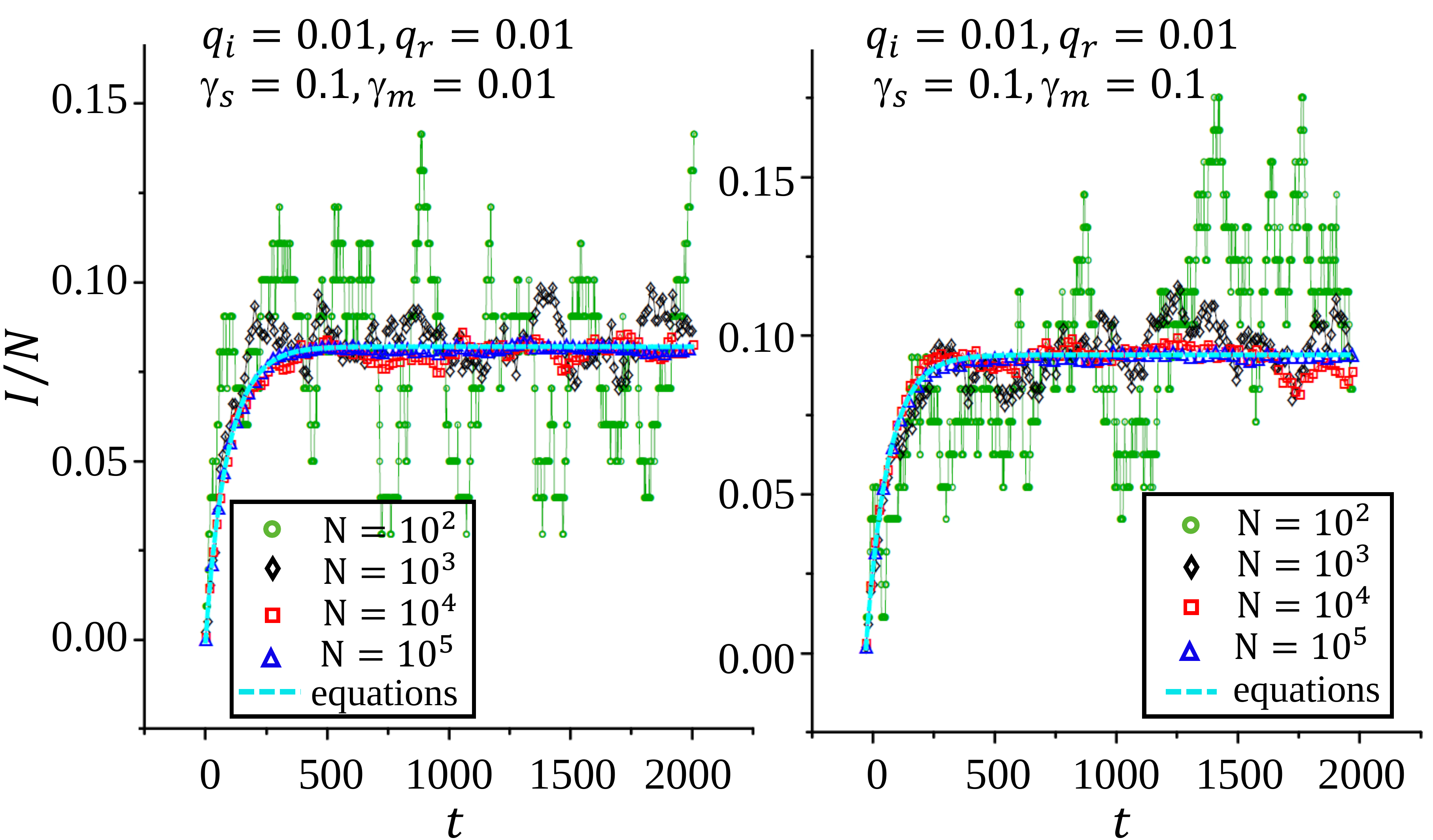}
\end{center}
\caption{(Color online) Fraction $I/N$ as a function of time $t$, for SIS model with a broadcast type infection mechanism.  Results for two sets of parameters and for different system sizes $N$ are shown.} \label{fig:bd02}
\end{figure}

We next consider the broadcast SIS mechanism, i.e. every susceptible who enters the group will be infected at a constant rate $q_i$. All the infected individuals have a recovery rate $q_r$ at which they become susceptible again. Interestingly, we note that this process is analogous to spintronics in condensed matter physics: When an electric current consisting of unpolarized electrons enters a spintronic device, such as a spin-valve transistor, the output current will be spin-polarized. The spin-polarized conducting electrons may be thought of as infected agents. The spin-polarized electrons will naturally tend to forget their polarization over time, e.g., by scattering (decoherence) or noise effects, and hence recover to become susceptible again (i.e. unpolarized). Any outbreak in the system may be described by the following equations in the mean-field limit:
\begin{eqnarray}
 \frac{{dS_g }}{{dt}}&=&  - q_i S_g  + q_r I_g  - p_l (S_g  - q_i S_g  + q_r I_g )\nonumber \\ 
&&+ p_j (S - S_g  + (I - I_g )q_r )\nonumber \\
 \frac{{dI_g }}{{dt}}&=& q_i S_g  - q_r I_g  - p_l (I_g  - q_r I_g  + q_i S_g )\nonumber \\ 
&&+ p_j (I - I_g )(1 - q_r )\nonumber \\
 \frac{{dS}}{{dt}}&=& - q_i S_g  + q_r I \nonumber \\
 \frac{{dI}}{{dt}}&=&q_i S_g  - q_r I
\end{eqnarray}
Some results are shown in Fig. \ref{fig:bd01}.  The simulation and numerical integration results generally agree with each other.
Compared to the numerical integration results, the simulation output fluctuates.  This is because the number of newly infected agents only depends on the number of susceptible in $G$.  It is therefore expected that the fluctuations will be smaller in a system of larger $N$, where simulation results will agree better with iteration results.  Recalling that the mean number of agents in $G$ is $N\gamma_{s}$, if we fix $\gamma_{s}$ and vary $N$ then the group size in $G$ will change.  Results for $I(t)/N$ with different values of $N$, are shown in Fig. \ref{fig:bd02}. The results for large $N$ show smaller fluctuations, as expected.

\begin{figure}
\begin{center}
\includegraphics[width=0.95\linewidth]{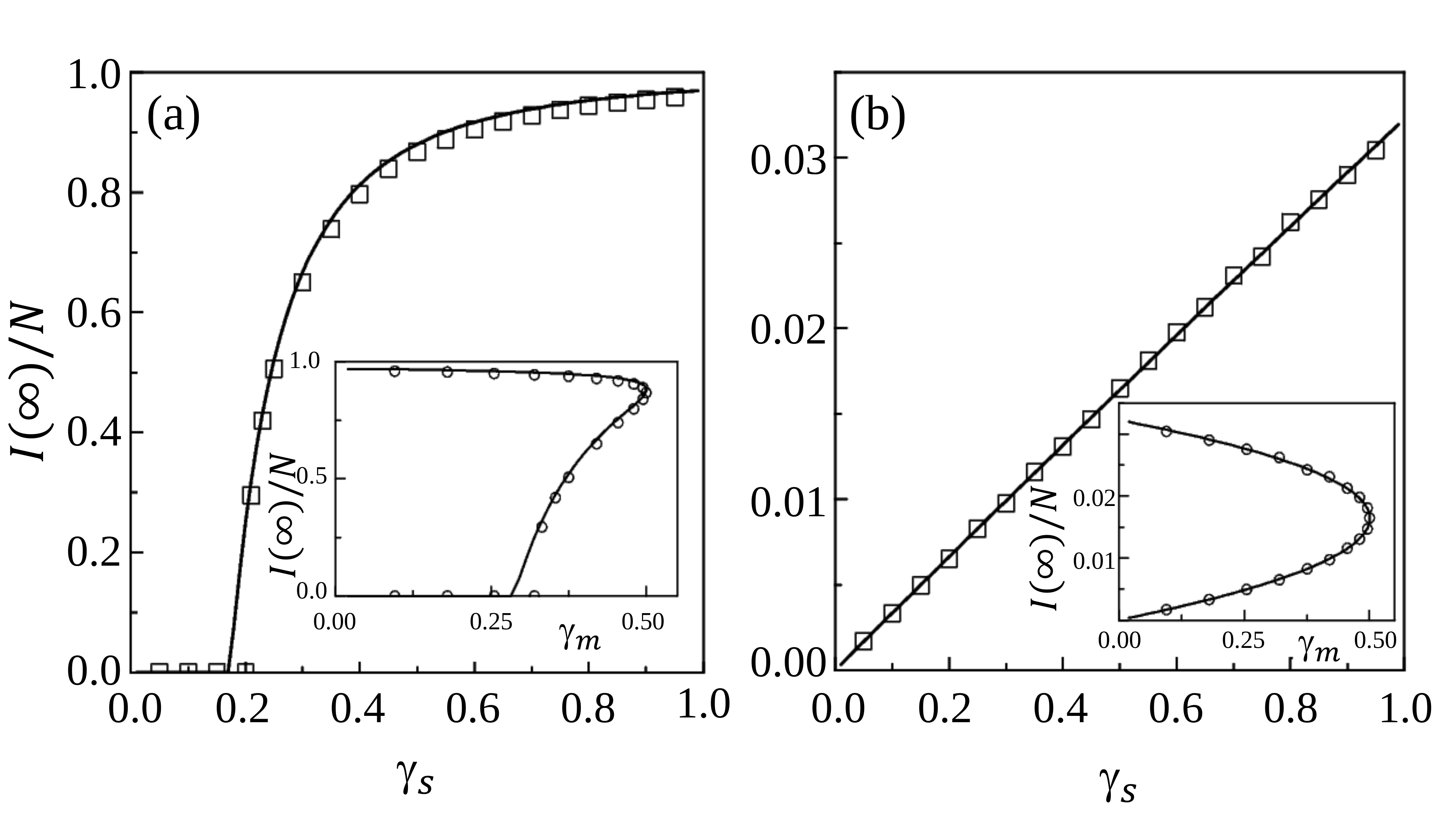}
\end{center}
\caption{Fraction $I(\infty)/N$ as a function of $\gamma_s$ (main panel). Insets show dependence of $I(\infty)/N$ on $\gamma_m$. Parameters are $N=1000$, $q_i=0.0005$, $q_r=0.015$. The special condition $p_j+p_l=1$ is satisfied, hence $\gamma_{m} = 2 \gamma_{s} (1 - \gamma_{s})$. Results shown for (a) person-to-person case and (b) broadcast case. Lines are analytic results from integrating the differential equations, while symbols are simulation results.} 
\label{fig:eq}
\end{figure}

The steady state $S(\infty)$ is given by
\begin{equation}
S({\infty}) = N + \frac{q_i k_1}{q_r k_2},
\end{equation}
where $k_{1}$ and $k_2$ are defined as
\begin{eqnarray}
k_1&=&Np_j q_r  + N\gamma _s q_r^2  -
N\gamma _s p_j q_r^2  - N\gamma _s p_l q_r^2 \\
k_2&=& - p_j q_i
- p_j q_r  - p_l q_r  - q_i q_r  + p_j q_i q_r  \nonumber\\&&+ p_l q_i q_r  -
q_r^2  + p_j q_r^2  + p_l q_r^2.
\end{eqnarray}
For the special condition $p_j+p_l=1$, we get
\begin{equation}
S({\infty}) = \frac{{Nq_r }}{{p_j q_i  + q_r }}
\end{equation}
and hence the fraction of infecteds is given by
\begin{equation}
\frac{I({\infty})}{N} =1- \frac{{q_r }}{{p_j q_i  + q_r
}}=\frac{{p_j q_i }}{{p_j q_i  + q_r }}\ .
\end{equation}
As for the SIR case under this same special condition, the equations take on the form of those for an effective well-mixed SIS system with effective parameters, that can be easily solved for $S(t)$. The solution is given by
\begin{equation}
S(t) = \frac{{Nq_r }}{{q_i \gamma _s  + q_r }} + Ce^{ - (q_i
\gamma _s  + q_r )t}\ .
\end{equation}
If the initial condition is $S(0)=N_0$, the constant $C$ is given by
\begin{equation}
C = N_0  - \frac{{Nq_r }}{{q_i \gamma _s  + q_r }}.
\end{equation}
Figure \ref{fig:eq} shows the results for how $I$ depends on $\gamma_s$ (which determines the group size in $G$) and $\gamma_{m}$ (which determines the mobility through $G$) for both the person-to-person case and broadcast case.  


\section{Summary}
We have presented and analyzed a simple but highly non-trivial model of co-existing mobility and infection dynamics. The model considers an SIR or SIS process for people transiting and revisiting a popular space $G$, with person-to-person and broadcast infection mechanisms. Our model can be solved through simulation, or by integrating a set of dynamical equations. Varying the mobility (i.e. agents entering and leaving the space $G$) and the infection probability has a significant impact on the overall infection profile even when both the mean group size in $G$ and the ratio of the infection and recovery probabilities are kept constant. The addition of a dynamical component through $G$, as compared to traditional well-mixed models, generates a far wider range of infection profiles and allows us to capture features observed recently in social outbreak phenomena. Our results reveal a highly non-linear dependence on mobility that generates the counter-intuitive prediction that by increasing the flow of individuals through a region of contagion, the infection's severity can be decreased. A special case arises for certain values of the parameters ($p_j + p_l = 1$) under which the system can be represented by an effective well-mixed model in which the group can be thought of as re-organized randomly in every time step. We also presented results for a broadcast-type infection mechanism and compared these to the person-to-person infection mechanism. We identified a range of infection probabilities where the two are comparable. More generally, we found that the wide variety of infection profiles that emerged from the time-dependent interplay between mobility and infection dynamics, was representative of recent real-world contagion phenomena in the social setting. 
 
\section{Acknowledgments} 
NFJ is grateful to the National Science Foundation (NSF) for funding under grant CNS1522693 and to the Air Force (AFOSR) for grant 16RT0367. The views and conclusions contained herein are solely those of the authors and do not represent official policies or endorsements by any of the entities named in this paper.



\begin{thebibliography}{99}
\bibitem{flu} D.G. McNeil, {\em Containing flu is not feasible, specialists say}, New York Times, Front Page, Thursday, April 30, 2009.
\bibitem{manore14} C.A. Manore, K.S. Hickmann, J.M. Hyman and I.M. Foppa, {\em A network-patch methodology for adapting agent-based models for directly transmitted disease to mosquito-borne disease}, E-print arXiv:1405.2258 [q-bio.QM], 2014.
\bibitem{stanley1} L.D. Valdez, H.H. Rego, H.E. Stanley and L.A. Braunstein. {\em Predicting the extinction of Ebola spreading in Liberia due to mitigation strategies.} Scientific Reports \textbf{5}12172, 2015.
\bibitem{children1}  Report by The American Academy of Pediatrics (AAP), {\em Pandemic Influenza: Warning, Children At-Risk} (2007).
\bibitem{onnela} J. Fernandez-Gracia, J.P Onnela, M.L. Barnett, V.M. Eguiluz and N.A. Christakis. {\em Spread of pathogens in the patient transfer network of US hospitals}. E-print arXiv:1504.08343, 2015.
\bibitem{boyle} S. Lyall, {\em Susan Boyle, Unlikely Singer Is YouTube Sensation}, New York Times, Front Page, Friday, April 17, 2009
\bibitem{havlin} L. Feng, Y. Hu, B. Li, H.E. Stanley, S. Havlin and L.A. Braunstein. {\em Competing for Attention in Social Media under Information Overload Conditions.} PLoS ONE \textbf{10}, e0126090, 2015.
\bibitem{finance} F.E. Harmon, Emergency Order by U.S. Securities and Exchange Commission: {\em False rumors can lead to a loss of confidence in our markets}, Release Number 58166, July 15, 2008.
\bibitem{neil} N.F. Johnson, P. Jefferies, and P.M. Hui. {\em  Financial Market Complexity}, (Oxford University Press, 2003). 
\bibitem{sornette1} D. Sornette, F. Desch\^atres, T. Gilbert,  and Y. Ageon,  Phys. Rev. Lett. {\bf 93}, 22 (2004).
\bibitem{riley} R. Crane, and D. Sornette, Proc. Nat. Acad. Sci. {\bf 105}, 15649 (2008).
\bibitem{Keeling} M.J. Keeling, and P. Rohani. {\em Modeling Infectious Diseases in Humans and Animals} (Princeton University Press, 2007).
\bibitem{May} R.M. May, and A.L. Lloyd, Phys. Rev. E {\bf 64}, 066112 (2001).
\bibitem{previous} F. Ball, D. Mollison, and G. Scalia-Tomba, Ann. Appl. Probab. {\bf 7} 46 (1997).
\bibitem{Havlin2} J. Shao, S. Havlin, and H.E. Stanley, Phys. Rev. Lett. {\bf 103}, 018701 (2009).
\bibitem{Koopman} J.S. Koopman, in {\em Biological Networks} ed. F. Kepes (World Scientific, London, 2007).
\bibitem{Murray} J.D. Murray, {\em Mathematical Biology: I. An Introduction} 3rd Edition (Springer, New York, 2007) Ch. 10.
\bibitem{cvespignani} V. Colizza, and A. Vespignani, Phys. Rev. Lett. {\bf 99}, 148701 (2007).
\bibitem{colizza} E. Valdano, L. Ferreri, C. Poletto, and V. Colizza, Phys. Rev. X {\bf 5}, 021005 (2015).
\bibitem{dodds} D.J. Watts, R. Muhamad, D.C. Medina, and P.S. Dodds, Proc. Nat. Acad. Sci. {\bf 102}, 11157 (2005); P.S. Dodds and D.J. Watts, Phys. Rev. Lett. {\bf 92}, 218701 (2004).
\bibitem{blasius} T. Gross, C. Dommar,  and B. Blasius, Phys. Rev. Lett. {\bf 96}, 20 (2006).
\bibitem{schwartz}  L.B. Shaw, and I.B. Schwartz, Phys. Rev. E, {\bf 77}, 066101 (2008).
\bibitem{us} N.F. Johnson. Complexity in Human Conflict, in {\em Managing Complexity: Insights, Concepts, Applications} edited by Dirk Helbing (Springer, Berlin, 2008) p. 303.
\bibitem{watts} D.J. Watts, R. Muhamad, D.C. Medina and P.S. Dodds. {\em Multiscale, resurgent epidemics in a hierarchical metapopulation model.} Proceedings of the National Academy of Sciences. {\bf 102}, 11157, 2005.
\bibitem{stanley2} L.G. Alvarez Zuzek, H.E. Stanley and L.A. Braunstein. {\em Epidemic Model with Isolation in Multilayer Networks.} Scientific Reports. {\bf 5}, 12151, 2015.

\bibitem{3} A. Stopczynski, A.S. Pentland and A. Lehmann. {\em Physical Proximity and Spreading in Dynamic Social Networks.} E-print arXiv:1509.06530v1 [physics.soc-ph], 2015.

\bibitem{murase} Y. Murase, J. Torok, H. Jo, K. Kaski and J. Kertesz. {\em Multilayer weighted social network model.} Physical Review E {\bf 90}, 052810, 2014.

\bibitem{vesp} M. Karsai, N. Perra and A. Vespignani. {\em Time varying networks and the weakness of strong ties.} Scientific Reports {\bf 4}, 4001, 2014.

\bibitem{scholtes14} I. Scholtes, N. Wider, R. Pfitzner, A. Garas, C.J. Tessone and F. Schweitzer. Nature Communications {\bf 5}, 5024, 2014.

\bibitem{barrat} A. Barrat, M. Barthelemy and A. Vespignani. {\em Dynamical Processes on Complex Networks}. Cambridge University Press, 2008.

\bibitem{baron} S. Liu, A. Baronchelli and N. Perra {\em Contagion dynamics in time-varying metapopulation networks.} Physical Review E {\bf 87}, 032805, 2013.

\bibitem{vesp2} R. Pastor-Santorras, C. Castellano and A. Vespignani. {\em Epidemic process in complex networks.} Reviews of Modern Physics. {\bf 87}, 925, 2015.


\bibitem{kaski} V. Palchykov, K. Kaski and J. Kertesz. {\em Transmission of cultural traits in layered ego-centric networks.} Condensed Matter Physics. {\bf 17}, 33802, 2014.

\bibitem{ker} H. Jo, J. Perotti, K. Kaski and J. Kertesz. {\em Analytically Solvable Model of Spreading Dynamics with Non-Poissonian Processes.} Physical Review X. {\bf 4}, 011041, 2014.

\bibitem{perra} M.V. Tomasello, N. Perra, C.J. Tessone, M. Karsai and F. Schweitzer. {\em The role of endogenous and exogenous mechanisms in the formation of R$\&$D networks.} Scientific Reports {\bf 4}, 5679, 2014.

\bibitem{10} T. Aoki, L.E.C. Rocha and T. Gross. {\em Temporal and structural heterogeneities emerging in adaptive temporal networks.} E-print arXiv:1510.00217v1 [physics.soc-ph], 2015.

\bibitem{5} P. Holme and J. Saramaki. {\em Temporal networks.} Physics Reports {\bf 519}, 97, 2012.

\bibitem{11more} E. Valdano, L. Ferreri, C. Poletto and V. Colizza. {\em Analytical Computation of the Epidemic Threshold on Temporal Networks.} Phys. Rev. X. {\bf 5}, 21005, 2015.

\bibitem{vesp09} P. Bajardi1, C. Poletto, D. Balcan, H. Hu, B. Goncalves, J.J. Ramasco, D. Paolotti, N. Perra, M. Tizzoni, W. Van den Broeck, V. Colizza and A. Vespignani. {\em Modeling vaccination campaigns and the Fall/Winter 2009 activity of the new A(H1N1) influenza in the Northern Hemisphere.} Emerging Health Threats Journal. {\bf 2}, e11, 2009.

\bibitem{11} P. Bajardi, A. Barrat, L. Savini and V. Colizza. {\em Optimizing surveillance for livestock disease spreading through animal movements.} J. of Royal Soc. Interface. {\bf 9}, 2814, 2012.

\bibitem{gonc} N. Perra, D. Balcan, B. Goncalves and A. Vespignani. {\em Towards a characterization of behavior-disease models.} PLoS ONE {\bf 6}, e23084, 2011.

\bibitem{estrada} E. Estrada. {\em The Structure of Complex Networks: Theory and Applications.} Oxford University Press, 2011.

\bibitem{Palla} G. Palla, A.L. Barabasi, and T. Vicsek, Nature {\bf 446}, 664 (2007).
\bibitem{us2} Z. Zhao, J.P. Calderon, C. Xu, G. Zhao, D. Fenn, D. Sornette,  R. Crane, P.M. Hui, N.F. Johnson. Effect of social group dynamics on contagion. Physical Review E {\bf 81}, 056107 (2010).
\bibitem{unrest} P. Manrique, H. Qi, A. Morgenstern, N. Velasquez, T. Lu, N.F. Johnson. 
Context Matters: Improving the Uses of Big Data for Forecasting Civil Unrest. IEEE Intelligence and Security Informatics, 169-172 (2013). ISBN 978-1-4673-6214-62013 

\bibitem{stepoanova11} E. Stepanova. The Role of Information Communication Technologies in the “Arab Spring”. PONARS Eurasia Policy Memo No. 159 (2011).

\bibitem{IARPA} J. Matheny. Test and evaluation in ACE and OSI IARPA  (2013). Available at \noindent ${\rm www.semanticommunity.info/@api/deki/files/21696/3}$
\noindent ${-ACE\_and\_OSI\_NIST\_Brief.pdf}$
\bibitem{Karsai2} M. Karsai, K. Kaski, A-L. Barabasi, J. Kertesz. Universal features of correlated bursty behavior. Scientific Reports {\bf 2}, 397 doi:10.1038/srep00397 (2012)




\end{thebibliography}
\end{document}